\documentclass[fleqn,usenatbib]{mnras}

\pdfoutput=1  

\usepackage{graphicx}            
\usepackage{amsmath}             
\usepackage{amssymb}             
\usepackage{gensymb}             
\usepackage{bm}                  
\usepackage{xspace}              
\usepackage{nicefrac}            
\usepackage{hyperref}            
\usepackage{xcolor}              
\usepackage{ifthen}
\usepackage{enumitem}
\usepackage{subfig}
\usepackage{suffix}

\title[The Hubble constant from eight time-delay galaxy lenses]{The Hubble constant from eight time-delay galaxy lenses}

\newcommand{\nauthor}[2]{#2,$^{#1}$}
\newcommand{\nauthorl}[2]{#2$^{#1}$}
\newcommand{\email}[1]{\thanks{Email: #1}}
\newcommand{\naffiliation}[2]{$^{#1}$#2}

\newcommand{\Gyrs}{\ensuremath{\mathrm{Gyr}}\xspace}
\newcommand{\GeVmcube}{\ensuremath{\mathrm{GeVm^{-3}}}\xspace}
\newcommand{\Ho}{\ensuremath{H_0}}
\newcommand{\Hunits}{\ensuremath{\mathrm{km}\,\mathrm{s}^{-1}\,\mathrm{Mpc}^{-1}}}
\newcommand\aHz{\ensuremath{\mathrm{aHz}}\xspace}
\newcommand{\Hres}{\ensuremath{71.8^{+3.9}_{-3.3}\,\Hunits{}}}
\newcommand{\HaHzres}{\ensuremath{2.33^{+0.13}_{-0.11}\,\mathrm{aHz}}}
\newcommand{\invHres}{\ensuremath{13.7^{+0.7}_{-0.7}\,\Gyrs{}}}
\newcommand{\rhocritres}{\ensuremath{5.4^{+0.6}_{-0.5}\,\GeVmcube{}}}

\newcommand{\seclbl}[1]{\label{sec:#1}}
\newcommand{\subseclbl}[1]{\label{subsec:#1}}
\newcommand{\figlbl}[1]{\label{fig:#1}}
\newcommand{\tablbl}[1]{\label{tab:#1}}
\newcommand{\eqlbl}[1]{\label{eq:#1}}

\newcommand{\secref}[1]{Section~\ref{sec:#1}}
\WithSuffix\newcommand\secref*[1]{Section~\ref{sec:#1}}
\newcommand{\subsecref}[1]{Subsection~\ref{subsec:#1}}
\WithSuffix\newcommand\subsecref*[1]{\ref{subsec:#1}}
\newcommand{\figref}[1]{Figure~\ref{fig:#1}}
\WithSuffix\newcommand\figref*[1]{\ref{fig:#1}}
\newcommand{\tabref}[1]{Table~\ref{tab:#1}}
\WithSuffix\newcommand\tabref*[1]{\ref{tab:#1}}
\renewcommand{\eqref}[1]{Eq.~(\ref{eq:#1})}
\WithSuffix\newcommand\eqref*[1]{(\ref{eq:#1})}

\newcommand*{\home}{./}%

\newcommand{\tref}[1]{\hyperlink{tr#1}{(#1)}}
\newcommand{\tlink}[1]{\hypertarget{tr#1}{(#1)}}

\def\qwidth{.24\textwidth}
\def\qheight{.24\textheight}

\author[Denzel et al.]{%
  \nauthor{1,2}{Philipp Denzel} \email{phdenzel@physik.uzh.ch}
  \nauthor{3}{Jonathan P. Coles} 
  \nauthor{2,1}{Prasenjit Saha} 
  \nauthorl{4}{Liliya L. R. Williams} 
  \newauthor
  \\
  \naffiliation{1}{Institute for Computational Science, University of Zurich, 8057 Zurich, Switzerland} \\
  \naffiliation{2}{Physics Institute, University of Zurich, 8057 Zurich, Switzerland} \\
  \naffiliation{3}{Physik-Department T38, Technische Universit\"at M\"unchen, James-Franck-Str.\ 1, D-85748 Garching, Germany} \\
  \naffiliation{4}{School of Physics and Astronomy, University of Minnesota, 116 Church Street SE, Minneapolis, MN 55455, USA} \\
}

\date{}
\pubyear{}

\begin{document}

\label{firstpage}
\pagerange{\pageref{firstpage}--\pageref{lastpage}}

\maketitle

\begin{abstract}
  \noindent 
  We present a determination of the Hubble constant from the joint, free-form analysis of 8 strongly, quadruply lensing systems.  
  In the concordance cosmology, we find $\Ho{} = \Hres{}$ with a precision of $4.97\%$.
  This is in agreement with the latest measurements from Supernovae Type Ia and Planck observations of the cosmic microwave background.  
  Our precision is lower compared to these and other recent time-delay cosmography determinations, because our modelling strategies reflect the systematic uncertainties of lensing degeneracies.
  We furthermore are able to find reasonable lensed image reconstructions by constraining to either value of $\Ho$ from local and early Universe measurements.
  This leads us to conclude that current lensing constraints on $\Ho$ are not strong enough to break the ``Hubble tension'' problem of cosmology.
\end{abstract}


\section{Introduction}\seclbl{td:intro}

In the flat $\Lambda$-cold dark matter model of cosmology ($\Lambda$CDM), the
rate at which the Universe expands on large scales is
\begin{equation}\eqlbl{td:adot}
  \frac{da}{dt} = \Ho \left( \frac{\Omega_m}a + \frac{\Omega_r}{a^2} +
  \Omega_\Lambda a^2 \right)^{1/2}
\end{equation}
where $a$ is the scale factor and $t$ is the cosmic time.  The Hubble
constant \Ho{} is defined as the $\dot{a}/a$ at the current epoch and
sets the overall scale.  The fractional contribution of the
non-relativistic and relativistic mass-energy and dark energy
components is captured by the $\Omega_i=\rho_i/\rho_c$
parameters, which are normalized by
\begin{equation}\eqlbl{td:rho_crit}
    \rho_{c} = \frac{3\Ho^2}{8\pi G}
\end{equation}
the cosmological critical density.  Local variations
(galaxies) around the mean density arise from an initial fluctuation
spectrum described by further cosmological parameters.  This model
\citep[for a `skeptic's guide' see][]{Scott18} succeeds in describing
a multitude of phenomena including the accelerating rate of expansion,
the statistics of fluctuations both in the local Universe and in the
cosmic microwave background (CMB), and the abundances of the light
elements.  Yet despite an increasing number of successes with
measurements of unprecedented precision, some unsolved puzzles remain.
Among these is the tension in the values of \Ho{} from standard
candles \citep[in particular, $74.0 \pm 1.4$\,\Hunits\ from the SH0ES
  (Supernovae \Ho{} for the Equation of State) project by][]{Riess19}
and from the CMB \citep[most recently $67.4 \pm 0.5$\,\Hunits\ from
  the Planck collaboration][]{Planck18b}.

These measurements (sometimes called `late' and `early') represent two
fundamentally different strategies for measuring cosmological parameters, and
involve completely different physical processes.  
The first of these involves a redshift-distance relation, whereby one measures
how the comoving distance
\begin{equation}\eqlbl{td:comovD}
  r(a) = c \int \frac{dt}{a(t)}
\end{equation}
(or some variant of it) depends on redshift $z=1/a-1$.  Standard-candle methods,
gravitational-lensing time delays, and anticipated methods using
gravitational-wave sources all use redshift-distance relations.  In contrast,
measurements of cosmological parameters from the CMB or from baryon acoustic
oscillations use a different strategy, where the main observable is the angular
power spectrum on the sky of acoustic oscillations in the Universe from epochs
when structure growth was linear.  The angular scale of the largest features is
set by the apparent size of the horizon $\theta_h(a)$ at the relevant redshift
where
\begin{equation}\eqlbl{td:horizon}
  \theta_h^{-1}(a) = \frac{r(0)}{r(a)} - 1.
\end{equation}
There is no explicit redshift-distance relation involved.  Instead,
\Ho{} is inferred through the effect of the component densities
$\propto\Ho^2/G$ on the acoustic oscillations.  Comparing the \Ho{}
values from these completely different physical processes is an
important test of the $\Lambda$CDM paradigm.  If the `Hubble tension'
is confirmed as a discrepancy, many alternative cosmological theories
will need to be considered \citep[see e.g.][]{Knox20}.

For the redshift-distance relation, thermonuclear supernovae (SNeIa) as standard
candles have been the leading method for some time \citep{Sandage06, Freedman12,
Riess16, Riess18, Riess19}.  The uncertainty in this technique is mainly that
the intrinsic brightness of SNeIa is difficult to determine, and requires a
`distance ladder' for calibration from other distance measurements in the local
Universe \citep{Pietrzyski19,Reid19,Freedman19}.  Distance measurement using
gravitational waves has only recently become feasible \citep{LIGOH0} and is
especially interesting because no separate calibration is required.  The period
and period derivative (chirp) of a gravitational-wave binary give $1+z$ times
the orbital energy of the binary.  The two gravitational-wave polarisations have
amplitudes of orbital energy divided by distance times inclination-dependent
factors.  If both polarisations are measured, inclination and distance both get
measured.  From acoustic oscillations, the CMB results from Planck
\citep{Planck18b} and earlier from WMAP \citep[Wilkinson Microwave Anisotropy
Probe;][]{WMAP9yr} are the best known, but there are also several measurements
of cosmological parameters using a combination of galaxy clustering, weak
lensing, baryonic acoustic oscillations, and primordial nucleosynthesis
\citep{DES1a, DES1b, Alsing16, Hildebrandt16, SDSS3BAO}.

Gravitational-lensing time delays present another form of redshift-distance
relation, involving multiple distances within one system.  As lensing time
delays are the subject of the present paper, we introduce the basic equation
here, in the variational formulation following \cite{Blandford86} with small
changes of notation.  Consider a virtual light ray originating at a source at
$\bm\beta$ on the sky, and at distance $D_S$ from the observer.  In front of
the source, at redshift $z_L$ and distance $D_L$ from the observer, lies a
gravitational lens consisting of a thin mass distribution $\Sigma(\bm\theta)$.
The virtual ray gets deflected at the lens and reaches the observer from
$\bm\theta$ on the sky.  The arrival-time surface $t(\bm\theta)$ of this
virtual light ray is
\begin{equation}\eqlbl{td:arrivfull}
  \frac{t(\bm\theta)}{(1+z_L)} =
        \frac{D_LD_S}{2cD_{LS}} (\bm\theta-\bm\beta)^2
       - \frac{8\pi G}{c^3} \nabla^{-2} \Sigma(\bm\theta)
\end{equation}
where $D_{LS}$ is the distance from the lens to the source.  Although these are
all angular-diameter distances, and hence $D_{LS}\neq D_S-D_L$, the distances
are still strictly proportional to $c/\Ho$.  That is, $\Ho^{-1}$ sets the scale
of \eqref{td:arrivfull}.  Real light rays correspond to $\nabla t(\bm\theta)=0$,
namely minima, saddle points, and maxima of $t(\bm\theta)$, and these are the
locations of multiple images.  If the source varies in time, the differences (or
time delays) between $t(\bm\theta_i)$ at multiple images can be measured.  For
variable sources such as quasars the differences in arrival times are usually of
the order several days to a year, but if the lens systems are highly symmetric
or have merging triplets, their delays can be well under an hour.  With accurate
measurements of the time delays, it is possible to determine a time scale which
is proportional to $\Ho^{-1}$.  This makes strong gravitational lenses excellent
cosmological probes because they enable a determination of \Ho{} completely
independent of the cosmic distance ladder.  Lensing time delays have therefore
been the subject of many observational campaigns, the most recent results
reported in \cite{Millon20a} and \cite{Millon20b}.

The early theoretical work \citep{Refsdal64,Refsdal66} considered
point-like sources and lenses.  But as soon as lenses were discovered,
it became clear that extended mass distributions $\Sigma(\bm\theta)$ and
extended sources would need modelling.  The resulting model-dependence
of inferences was noted already in the first paper modelling lensing
data \citep{Young80} and has been explored in many later works
\cite[e.g.][]{Saha06,SchneiderSluse14,Wagner18,Denzel20}.  If there
are many lensed sources at different redshifts, each lensed into
multiple images, $\Sigma(\bm\theta)$ will be well constrained by them.
\cite{Ghosh20} estimate that if there are 1000 lensed images, a single
precise time-delay measurement would measure \Ho{} to sub-percent
accuracy, and that this may be feasible with JWST observations of
cluster lenses.  Galaxy lenses, however, rarely have more than one
source lensed into four images, so the best strategy is to combine
many lenses. That said, multiple-source lens systems such as the
``jackpot'' double Einstein ring, for which just recently a third source
has been reported by \cite{Collett20}, have been found to reduce
degeneracies substantially and should therefore reduce uncertainties
on \Ho{} inferences \citep{Gavazzi08} if time-delay measurements in
such systems become available.  \cite{Paraficz14} report a 10\%
uncertainty on \Ho{} using 18 time-delay lenses.  Smaller
uncertainties on \Ho{} are possible---2.5\% from six lenses by the
H0LiCOW collaboration \citep[\Ho{} Lenses in COSMOGRAIL's
Wellspring;][]{Wong20} and 4\% uncertainty from a single lens
\citep{Shajib20}---if it is assumed that galaxy lenses follow certain
parametric forms.  \cite{Gilman20} demonstrate using mock observations
based on real lens configurations that perturbations from substructure
contribute an additional source of random uncertainty in the inferred
value of \Ho{}. If these substructures are fitted properly, they could
to a certain degree improve cosmographic inferences from single galaxy
lenses. Uncertainties on \Ho{} increase however if lensing
degeneracies are considered, but may be alleviated again if stellar
kinematics are considered, e.g. \citep{TDCOSMO4}. In a recent blind
test involving several research groups, the Time Delay Lens Modelling
Challenge \citep[TDLMC;][]{TDLMC2}, the currently achievable error
level in the recovery of simulated \Ho{} from up to 16 lenses was
found to be 6\%.

The Hubble constant is commonly expressed in units of \Hunits{}.  This
choice reflects Hubble's law by stating that the Hubble constant is
the recession speed of a target galaxy over its distance.  However, it
is arguably more natural to think of the Hubble constant as a
reciprocal time or frequency.  Of course, the Hubble time $\Ho^{-1}$
\emph{is already} in units of \Gyrs{} and proportional to the age of
the Universe.  In the current epoch, the Hubble parameter seems to be
close to steady and consequently, distances scale (nearly)
exponentially with $a \propto \mathrm{e}^{Ht}$ due to the increasingly
dominant dark-energy density component.  In this context, it is
interesting to express the Hubble constant in SI units as
\textit{attohertz} ($\mathrm{aHz} = 10^{-18}\,\mathrm{s}^{-1}$).
Since the Hubble tension has been a persistent problem, a change of
units also provides a new perspective on the issue and might promote
new ideas.  One should recall that the following are all equivalent:
%
    \begin{alignat*}{2}
        &\Ho      &&= 70.0 \, \Hunits = 2.27 \, \aHz{} \\
        &\Ho^{-1} &&= 14.0 \, \Gyrs{} \\
        \frac{3}{8\pi G}&\Ho^2 &&= 5.16 \, \GeVmcube{}
  \end{alignat*}
%

These new units of attohertz then convey that large-scale structures undergo
e-foldings with a frequency of roughly 2.3~aHz.  We can compare this current day
value to that of the very early Universe where 60 e-folds may have occurred in
$\sim$1~s~\citep{Allahverdi20}.

In this paper, we infer \Ho{} from 8 free-form time-delay lenses using the most
recent observational data available and more flexible modelling methods.
Additionally, we explore the effect of lensing degeneracies on \Ho{}, which is
the inherent limitation to all lensing observables \citep{Saha2000}. We
demonstrate that due to these degeneracies it is possible to find solutions
which fit values of \Ho{} measured through early and late-Universe probes.  Our
main findings are summarized in \figref{td:H0filtered}.

\begin{figure}
    \includegraphics[width=\hsize]{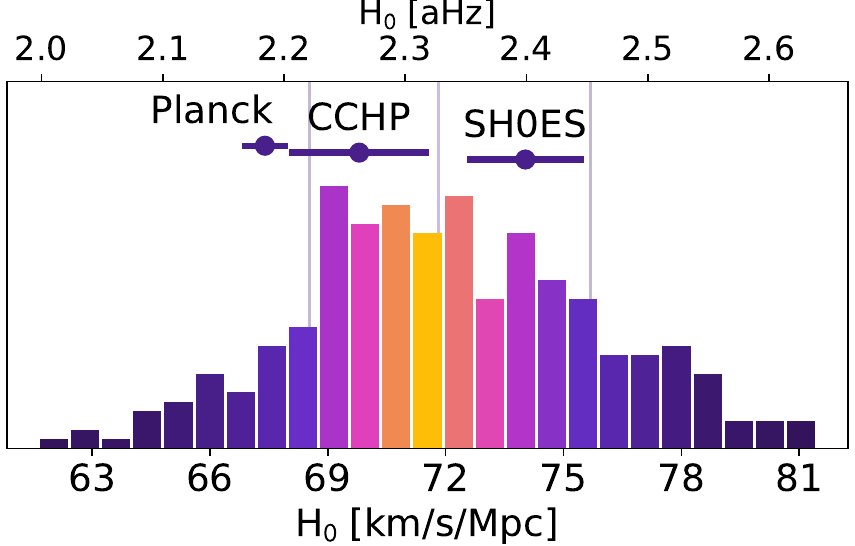}
    \caption{The main result: an ensemble of \Ho{} values inferred from the
      joint modelling of eight four-image time-delay lenses.  Vertical lines
      indicate the median 68 per-cent confidence range of $\Ho = \HaHzres{} =
      \Hres{}$.  To make the median furthermore easily discernible, the
      colouring of the histogram's bars corresponds to the cumulative
      probability centred around the median (yellow-magenta-blue-black goes from
      1 to 0).  Horizontal error bars indicate recent measurements from other
      methods for comparison: Planck \protect\citep{Planck18b},
      CCHP\protect\citep[the Carnegie-Chicago Hubble Program;][]{Freedman19},
      and SH0ES \protect\citep[the Supernovae \Ho{} for the Equation of
      State;][]{Riess19}.  A discussion of units can be found in the
      introduction.}
      \figlbl{td:H0filtered}
\end{figure}

In \secref{td:systems} below, we describe the eight lenses used in
this work: B1608+656, DESJ0408-5354, HE0435-1223, PG1115+080,
RXJ0911+0551, RXJ1131-1231, SDSSJ1004+4112, and WFIJ2033-4723.  For
each system we give an account of the research history, a short
description of the image configuration and point out any other special
features.  Next, in \secref{td:methods}, we describe the numerical
techniques and methods employed to analyse the lenses and determine
\Ho{}. We list precisely what information was used to model and
constrain the lens reconstructions.  Here, \subsecref{td:tdlmc}
describes the TDLMC in which our lens-modelling methods have been
initially tested for the purpose of inferring \Ho{}. In
\secref{td:results}, we present the detailed results of our work,
namely the lens models, and derived quantities such as
radially-averaged enclosed mass profiles, time-arrival surfaces,
synthetic images, and posterior distributions of \Ho, and moreover of
the Hubble time and critical density.  Relevant results from the TDLMC
are discussed at the end of \subsecref{td:H0}.  Finally in
\secref{td:conclusion}, we summarize the main findings and discuss
possible implications.

\section{The lens systems}\seclbl{td:systems}

  The light travel time of the individual images differ if strongly lensed
  systems lie at cosmological distances.  However, measuring time delays is a
  very time-consuming process and only a handful of systems are currently known
  with comparatively precise and robust values, meaning if multiple independent
  monitorings have been performed, they concur within their error margins.  Due
  to computational constraints and their strong presence in the literature, we
  limited our analysis to the following 8 quadruply imaging lenses shown in
  \figref{td:composites}.  In this section, we briefly describe the lenses and
  note the time delays with the image ordering provided in the respective
  literature.

  \begin{figure*}
    \centering
    \setlength{\tabcolsep}{3pt}
    \begin{tabular}{llll}
        \includegraphics[width=\qwidth]{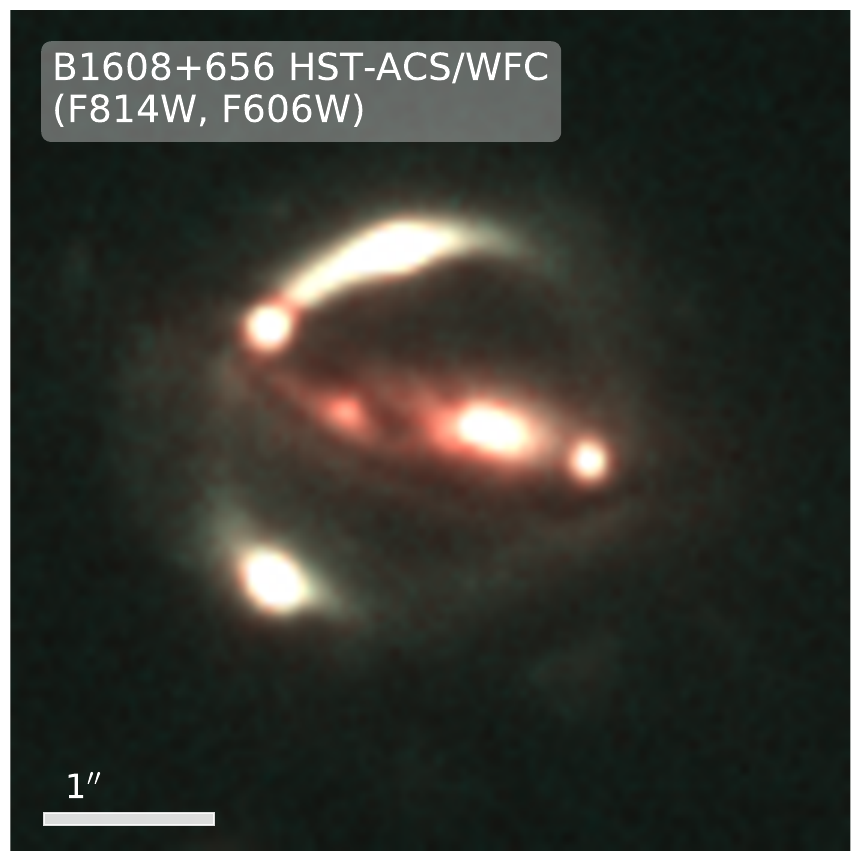} & \includegraphics[width=\qwidth]{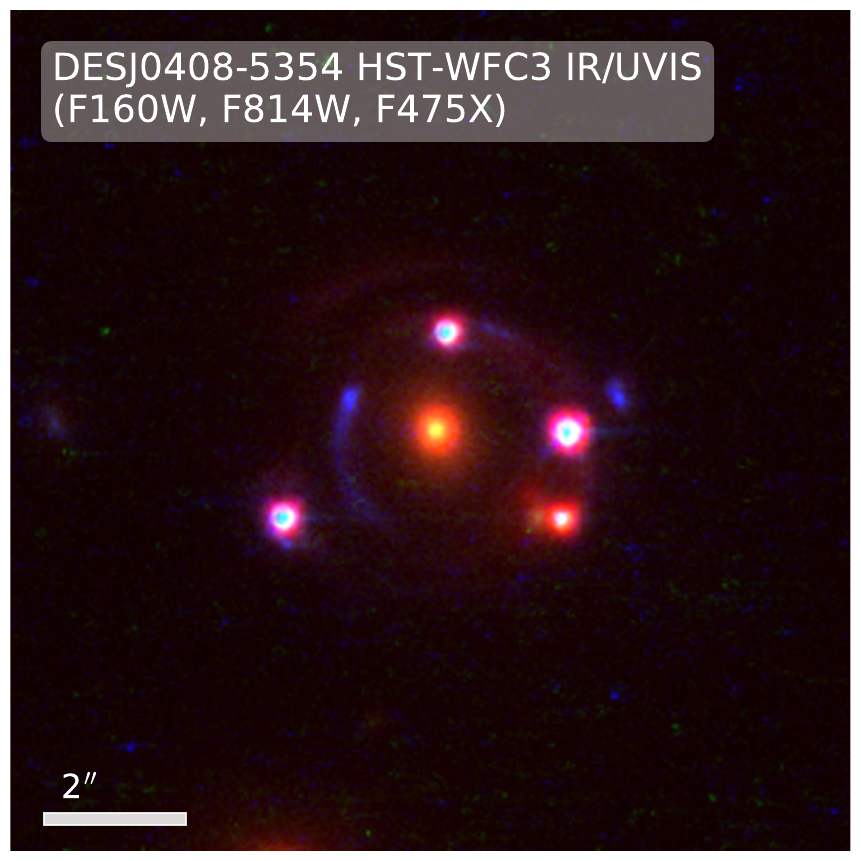} & \includegraphics[width=\qwidth]{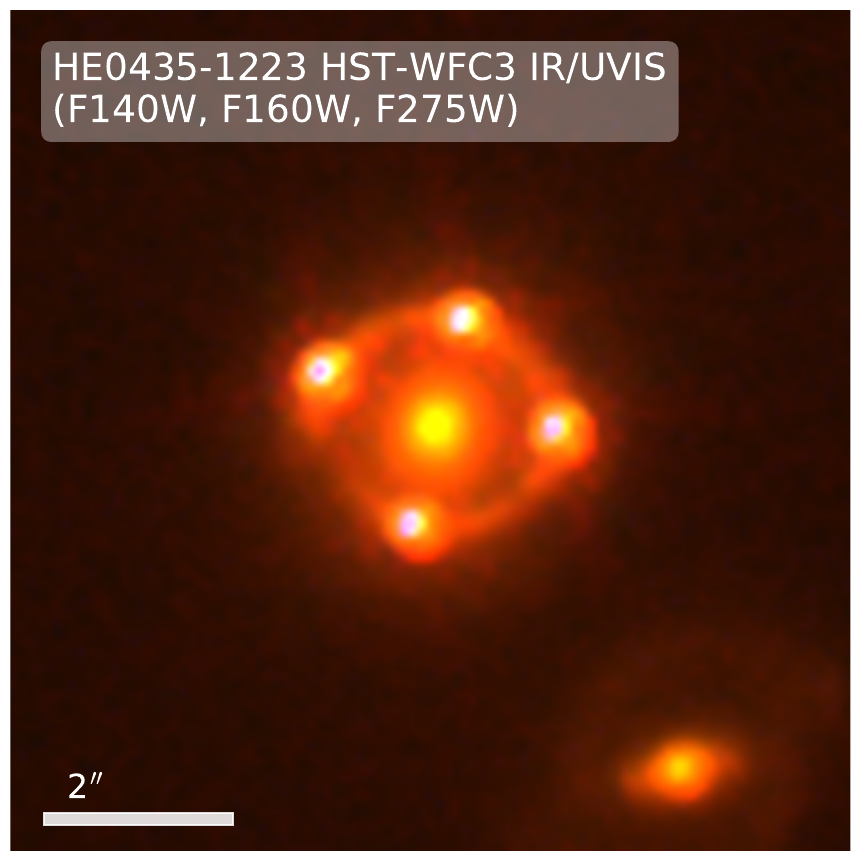} & \includegraphics[width=\qwidth]{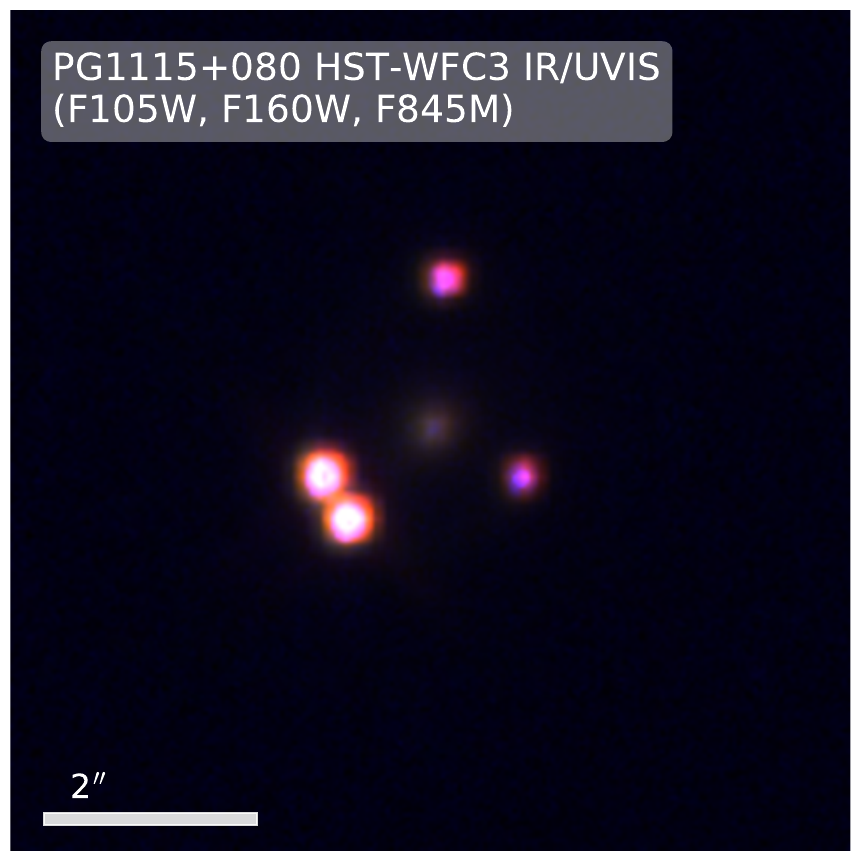}\\%
        \includegraphics[width=\qwidth]{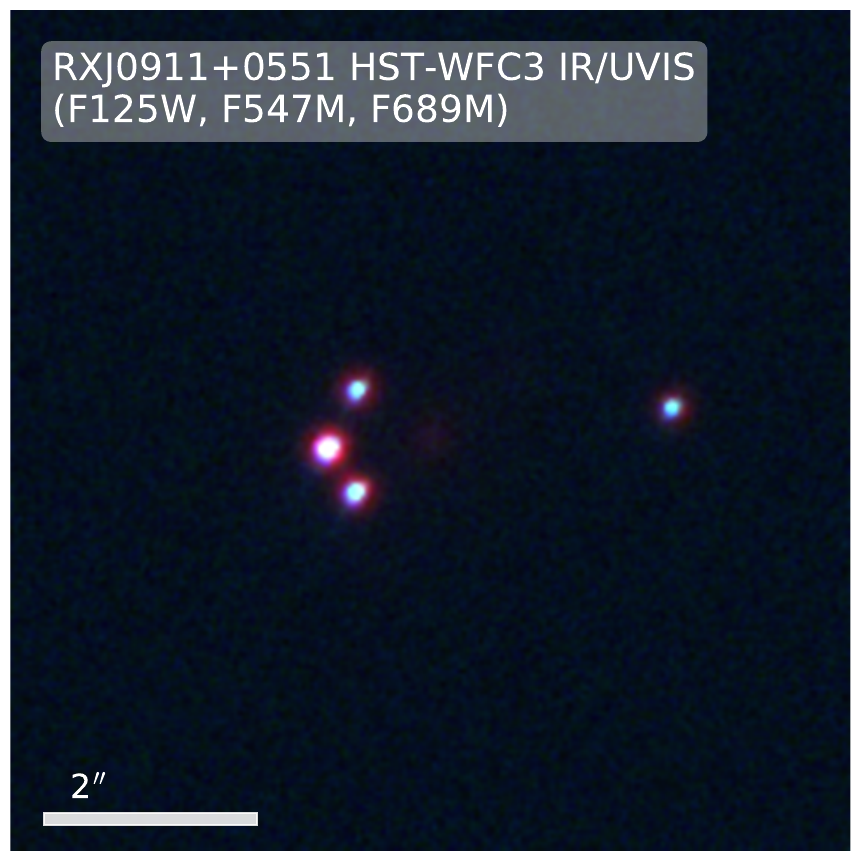} & \includegraphics[width=\qwidth]{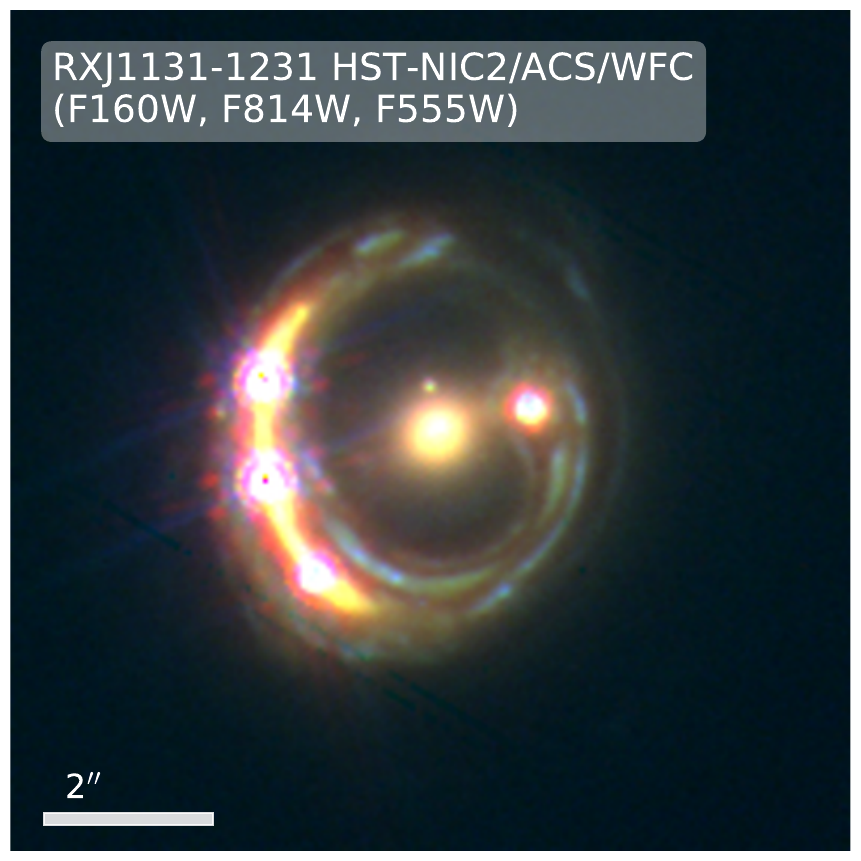} & \includegraphics[width=\qwidth]{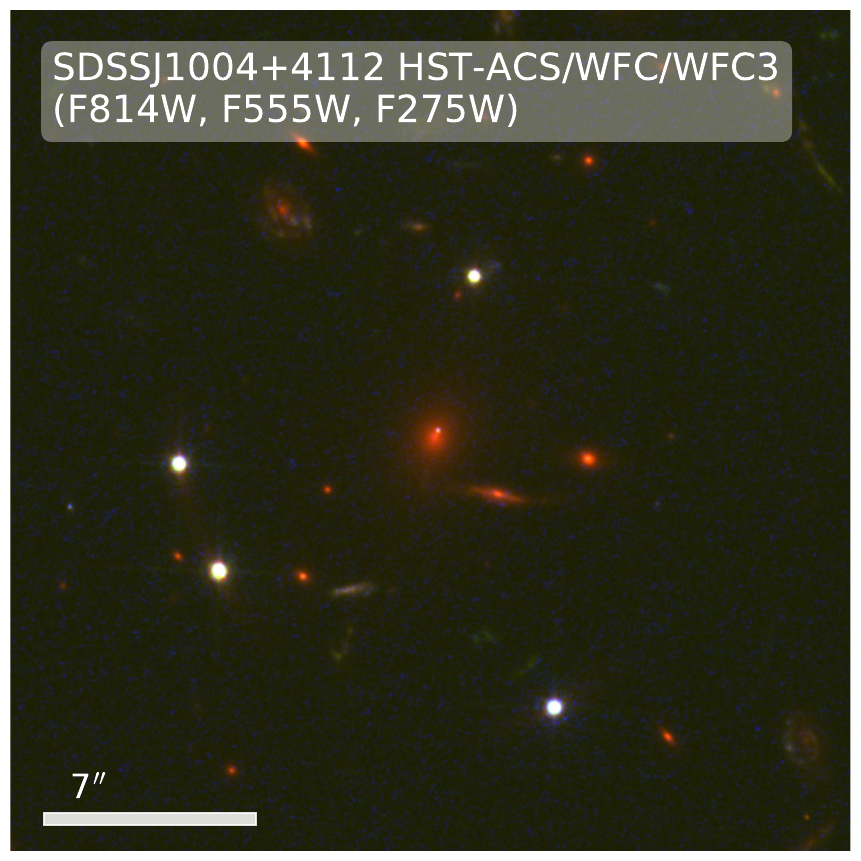} & \includegraphics[width=\qwidth]{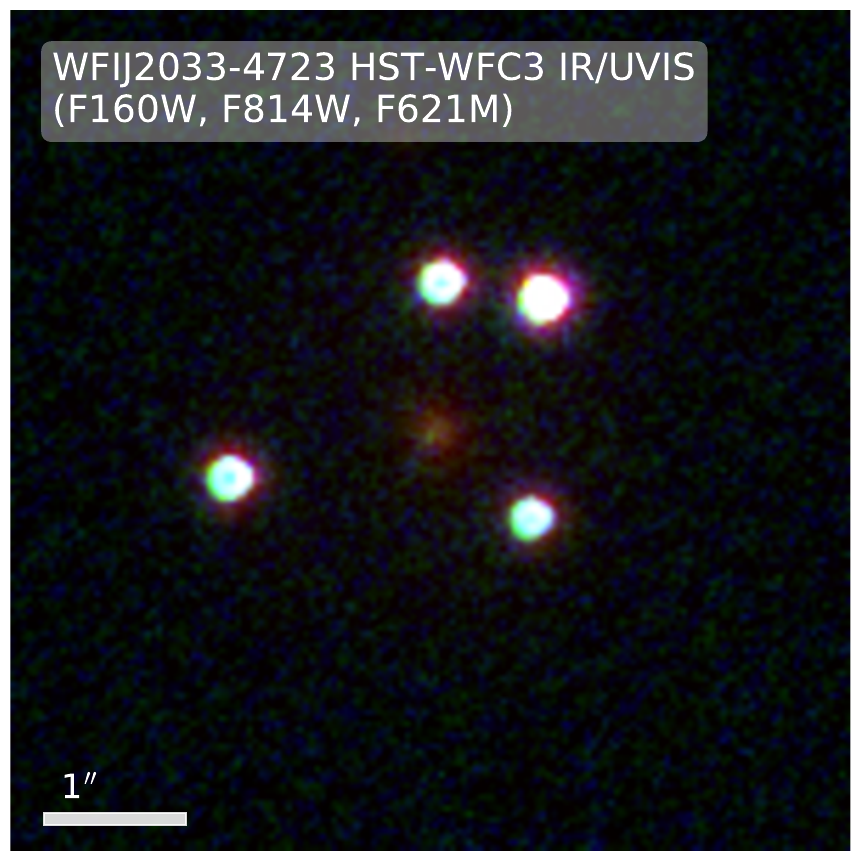}%
    \end{tabular}
    \caption{Composite HST images of the modelled lens systems centred on the
    main lens.  The image data were stacked using 2 or 3 bands in the
    UV/Visible, optical, or near-infrared with an individually modified method
    described by \protect\cite{Lupton04}.  The exact camera system and filters
    are listed in the label ordered by R, G, and B colour channels.  In all
    images north is up and east is left.  }\figlbl{td:composites}
\end{figure*}
  
  \subsection{B1608+656}\subseclbl{td:B1608} This system was discovered during
  the first phase of the Cosmic Lens All-Sky Survey (CLASS).  The system
  contains two lensing galaxies and clearly shows four quasar images on radio
  wavelengths, whereas its Einstein ring is better visible in the optical
  spectrum.  Further data collection yielded redshifts for the lens
  $z_{\text{l}} = 0.630$, the source $z_{\text{s}} = 1.394$, and hints that the
  lens consists of two merging galaxies \citep{B1608Myers95, B1608Fassnacht96}.
  After multiple seasons of monitoring \citep{B1608Fassnacht99,
  B1608Fassnacht01, B1608Fassnacht02} and several robustness tests
  \citep{Eulaers11, Holanda16} time delay measurements converged to
  $\Delta t_{\text{AB}} = 31.5^{+2.0}_{-1.0}$,
  $\Delta t_{\text{CB}} = 36.0^{+1.5}_{-1.5}$, and
  $\Delta t_{\text{DB}} = 77.0^{+2.0}_{-1.0}$ days with
  arrival-time order BACD.
  Since its discovery, many lens modellers have used the system for a
  determination of \Ho{} \citep{B1608Koopmans99, Williams2000, B1608Koopmans03,
  B1608Fassnacht05, B1608Suyu10, Wong20}.  
  
  \subsection{DESJ0408-5354}\subseclbl{td:DESJ0408}
  This system was found and confirmed only recently in the Dark Energy Survey
  (DES) Year 1 (Y1) data.  Subsequent spectroscopic observations using the
  Gemini South telescope confirmed a quasar in the source with redshift
  $z_{\text{s}} = 2.375$, and the central lens as an early-type galaxy with
  redshift $z_{\text{l}} = 0.597$ \citep{DESJ0408Lin17}.  Simultaneously,
  high-cadence photometric monitoring results by \citep{DESJ0408Courbin18}
  report time-delays measurements of 
  $\Delta t_{\text{AB}} = -112.1^{+2.1}_{-2.1}$, 
  $\Delta t_{\text{AD}} = -155.5^{+12.8}_{-12.8}$, and
  $\Delta t_{\text{BD}} = 42.4^{+17.6}_{-17.6}$ days 
  with arrival-time order ABCD.  
  Follow-up investigations revealed the presence of a second set of multiple
  images at different redshifts \citep{DESJ0408Agnello17}, which places it
  amongst the richest discoveries of lenses so far.  In depth analysis of this
  system by the STRong-lensing Insights into Dark Energy Survey collaboration
  \citep[STRIDES;][]{STRIDES} determined $\Ho = 74.2^{+2.7}_{-3.0}$
  \citep{Shajib20}.
  
  \subsection{HE0435-1223}\subseclbl{td:HE0435}
  \cite{HE0435Wisotzki02} reported the discovery of this system.  It has a
  lensed QSO at redshift $z_{\text{s}} = 1.689$ in a crossed image
  configuration.  The redshift $z_{\text{l}} = 0.454$ of the lensing galaxy was
  measured three years later by \cite{HE0435Morgan05}, and time delays were
  reported the following year by \cite{HE0435Kochanek06} as
  $\Delta t_{\text{AC}} = -2.1^{+0.78}_{-0.71}$, 
  $\Delta t_{\text{AD}} = -14.37^{+0.75}_{-0.82}$, and 
  $\Delta t_{\text{AB}} = -8.0^{+0.73}_{-0.82}$.
  The system has recently been the focus of many studies by the H0LiCOW
  collaboration which provided a spectroscopic survey \citep{HE0435Sluse17}, a
  mass model \citep{HE0435Wong16}, newly measured time delays
  \citep{HE0435Bonvin17}, and external convergence field models
  \citep{HE0435Tihhonova18}.  Moreover, \cite{HE0435Nierenberg17} investigated
  the lens' substructure with a WFC3 grism narrow-line survey.  Recently
  \cite{Millon20a} estimated the time delays as
  $\Delta t_{\text{AB}} = -9.0^{+0.8}_{-0.8}$, 
  $\Delta t_{\text{AC}} = -0.8^{+0.8}_{-0.7}$, 
  $\Delta t_{\text{AD}} = -13.8^{+0.8}_{-0.8}$, 
  $\Delta t_{\text{BC}} = 7.8^{+0.9}_{-0.9}$, 
  $\Delta t_{\text{BD}} = -5.4^{+0.9}_{-0.8}$, and 
  $\Delta t_{\text{CD}} = -13.2^{+0.8}_{-0.8}$ days
  with arrival-time order CADB.

  \subsection{PG1115+080}\subseclbl{td:PG1115} 
  The so-called \textit{triple quasar} is the second gravitational lens ever
  discovered \cite{PG1115Weymann80}.  Only after its naming, it was shown that
  the seemingly brightest image was---according to generic lens theory---in fact
  a blend of two separate images.  The elliptical lensing galaxy was detected
  only 7 years after its discovery by \cite{PG1115Christian87}.
  \cite{PG1115Kundic97} and \cite{PG1115Tonry98} independently measured the
  redshift of the lens to be $z_{\text{l}} = 0.311$.  The potential to constrain
  cosmological parameters was realized early and time delays were first measured
  by \cite{PG1115Schechter97}, and soon after improved by
  \cite{PG1115Barkana97}.  The latest time delays were reported by
  \citep{PG1115Bonvin18} as 
  $\Delta t_{\text{AB}} = 8.3^{+1.5}_{-1.6}$, 
  $\Delta t_{\text{AC}} = 9.9^{+1.1}_{-1.1}$, and 
  $\Delta t_{\text{BC}} = 18.8^{+1.6}_{-1.6}$ days 
  with arrival-time order CA$_{1}$A$_{2}$B.

  \subsection{RXJ0911+0551}\subseclbl{td:RXJ0911}
  \cite{RXJ0911Bade97} confirmed this system to be gravitationally lensing after
  follow-up observations of candidates from the ROSAT All-Sky Survey (RASS). It
  shows a complex image configuration with one image particularly far from the
  other three, requiring a large external shear component \cite{RXJ0911Burud98}.
  It was suspected that the origin of the high shear was a nearby cluster, which
  was finally discovered two years later by \cite{RXJ0911Kneib00}.
  \cite{RXJ0911Hjorth02} measured a time delay of $\Delta t_{\text{AB}} =
  146^{+8}_{-8}$ days with arrival-time order BA$_{1}$A$_{2}$A$_{3}$, where B is
  the distant and leading component image, and A$_i$ the combination of the
  other three.

  \subsection{RXJ1131-1231}\subseclbl{td:RXJ1131} The system has been
  serendipitously unveiled during polarimetric imaging of radio quasars by
  \cite{RXJ1131Sluse03}.  The redshifts of the lens and source were measured as
  $z_{\text{l}} = 0.295$ and $z_{\text{s}} = 0.658$ \citep{RXJ1131Sluse07}.  HST
  imaging shows a detailed Einstein ring from the host galaxy of the multiply
  imaged quasar.  Substructure-analyses by \cite{RXJ1131Claeskens06} and
  \cite{RXJ1131Suyu13} furthermore point out a satellite of the lensing galaxy,
  identifiable as a small, bright spot north of the lens. An extensive time
  delay estimation was obtained from a 9-year optical monitoring using three
  different numerical methods between all possible pairs of quasar images with
  arrival-time order BCAD by \cite[]{RXJ1131Tewes13}.  Due to the high quality
  of observational data available, the system has been subjected to modelling
  methods and cosmographic analyses in many previous works
  \cite[e.g.][]{Holanda16, RXJ1131Birrer16, RXJ1131Birrer17, Suyu17}.

  \subsection{SDSSJ1004+4112}\subseclbl{td:SDSSJ1004}
  With its multiple quasar images at a separation of $14.62\arcsec$, this system
  is the largest and rarest object in our set \citep{SDSSJ1004Inada03}.  As
  expected from such a giant, it was shown to be a cluster dominated by dark
  matter \citep{SDSSJ1004Oguri04, SDSSJ1004Williams04}.  Since time delays are
  of the order of $c^{-3}GM$ it is only natural to expect much higher delays for
  cluster lenses.  \cite{SDSSJ1004Oguri10} used mass models of the lensing
  galaxies to predict time delays of $\Delta t_{\text{AD}} \sim 1218$ days and
  \cite{SDSSJ1004Mohammed15} demonstrated how useful such a time delay
  measurement would be for constraining its substructure. After more than 3.5
  years of data collection the time delays were estimated as 
  $\Delta t_{\text{BA}} = 40.6^{+1.8}_{-1.8}$, 
  $\Delta t_{\text{CA}} = 821.6^{+2.1}_{-2.1}$, and 
  $\Delta t_{\text{AD}} > 1250$ days
  with arrival-time order CBAD \citep{SDSSJ1004Fohlmeister08,
  SDSSJ1004Fohlmeister07}.

  \subsection{WFIJ2033-4723}\subseclbl{td:WFIJ2033}
  Very similar to B1608+656 (see~\subsecref*{td:B1608}) and PG1115+080
  (see~\subsecref*{td:PG1115}), WFIJ2033-4723 is a quad in a fold configuration
  with two images almost blending.  It was discovered by \cite{WFI2033Morgan04}
  during an optical imaging ESO survey.  The quasar images are considerably
  brighter than the foreground lens galaxy whose redshift $z_{\text{l}} = 0.661$
  was measured by \cite{WFI2033Eigenbrod06}.  The redshift of the quasar images
  $z_{\text{s}} = 1.662$ were confirmed by \cite{Sluse12}.  The most recent time
  delay measurements report 
  $\Delta t_{\text{AB}} = 36.2^{+0.7}_{-0.8}$, 
  $\Delta t_{\text{AC}} = -23.3^{+1.2}_{-1.4}$, and 
  $\Delta t_{\text{BC}} = -59.4^{+1.3}_{-1.3}$ days
  with arrival-time order BA$_{1}$A$_{2}$C
  \citep{WFI2033Bonvin19}.

\newcommand\rpix{$\mathrm{r}_{\text{pix}}$}
\newcommand\spix{$\mathrm{s}_{\text{pix}}$}
\newcommand\GLASS{\textsc{GLASS}}

\section{Lens reconstruction method}\seclbl{td:methods}

The lenses were reconstructed using the free-form modelling code
{\GLASS}\footnote{GLASS: \url{http://ascl.net/1806.009}} \citep{GLASS} plus
recent developments by \cite{Denzel20}.  \GLASS\ has previously been used for
several studies of galaxy lenses \citep{GLASS,Kueng15,Bruderer16,Kueng18} and
cosmological parameters \citep{Lubini14} although not for \Ho, other than the
TDLMC.  However the related older code PixeLens\footnote{PixeLens:
\url{https://ascl.net/1102.007}} has been used to estimate \Ho{}
\citep{PixeLens,Paraficz14,Saha06c}.

\subsection{GLASS}

To formulate the method, we introduce the usual dimensionless density $\kappa$
and arrival time $\tau$
\begin{equation}\eqlbl{td:dimensionless}
  \begin{aligned}
      \Sigma(\bm\theta) &= \frac{c^3}{4\pi G\Ho} \frac{d_Ld_S}{d_{LS}} \,\kappa(\bm\theta) \\
      \Ho\,t(\bm\theta) &= (1+z_L) \frac{d_Ld_S}{d_{LS}} \,\tau(\bm\theta)
  \end{aligned}
\end{equation}
where $d_L$ is $D_L$ with the dimensional factor $c/\Ho$ taken out.
In terms of these, the arrival time becomes
\begin{equation}\eqlbl{td:arriv}
  \tau(\bm\theta) = \bm{\theta\cdot\beta} + {\textstyle\frac12}|\bm\theta|^2 - 2\nabla^{-2} \kappa(\bm\theta) \,.
\end{equation}

In free-form or pixellated lens reconstruction, the mass distribution
\begin{equation}\eqlbl{td:free-form-potential}
    \kappa(\bm{\theta}) = \sum \kappa_{n} Q(\bm{\theta}-\bm{\theta}_n)
\end{equation}
is represented as a collection of pixels where $Q(\bm{\theta}-\bm{\theta}_n)$
is a square (or other profile) and $\bm\theta_n$ is its centroid.  The
contribution $\nabla^{-2}Q(\bm{\theta})$ of a square to the arrival time can be
calculated analytically \citep{Abdelsalam98}.

The pixels $\kappa_n$ are arranged in concentric `pixel rings' centred on the
lensing galaxy.  The central pixel can be refined into sub-pixels in order to
better resolve steep cusps.  The $\kappa_n$ values are required to satisfy the
following prior inequality constraints.
\begin{enumerate}[leftmargin=*]
  \item All mass densities must be non-negative: $\kappa_{n} \geq 0$.
  \item In order to keep the mass distribution relatively smooth, no
    mass tile can exceed twice the average of its neighbors.
  \item The local density gradient $\nabla\kappa$ should point
    within $\alpha = 60^{\circ}$ of radially inwards:
    $$ \mathbf{R}\nabla\kappa \geq 0, \qquad \mathbf{R}^\intercal\nabla\kappa \geq 0$$ 
      where
      $\mathbf{R}=\mathbf{R}(\alpha)$ is a rotation.
  \item The average density $\langle\kappa\rangle_{i}$ of mass within a ring~$i$
    at radius $R_i$ must have a profile that is steeper than $R^{-s}$:
    $$ R_i^{-s}\langle\kappa\rangle_{i} - R_{i+1}^{-s}\langle\kappa\rangle_{i+1}
    \geq 0$$ This still allows for twisting iso-density contours and
    significantly varying ellipticities with radius. In this work we
    set $s=0$, requiring only that the circularly averaged surface mass
    density does not increase with radius, which is rather
    conservative.
\end{enumerate}

Additionally, since the pixels cover only a relatively small, finite surface,
external shear from, e.g., a nearby galaxy, can be added to \eqref{td:arriv}, as
a two-component shear
\begin{equation}\eqlbl{td:twocompshear}
  \gamma_1(\theta_x^2 - \theta_y^2) + 2\gamma_2\,\theta_x\theta_y
\end{equation}
where $(\gamma_1,\gamma_2)$ are constant shear components.  Furthermore,
neighbouring galaxies can be modelled as point masses (or other distributions)
whose contributions also add to \eqref{td:arriv}.

Multiple-image data from observations further constrain the mass distribution:
\begin{enumerate}[leftmargin=*]
\item The arrival time must be stationary at the observed image locations.  That
  is, for all image locations $\bm\theta_i$ we require
  $\nabla\tau(\bm\theta_i)=0$.  For one image, such an equation simply relates
  the unknown source position to the image position, so it does not constrain
  the mass distribution.  Additional images from the same source do, however,
  constrain the mass distribution, because the source position has already been
  determined by the first image.
\item The elements of the inverse magnification tensor
  $\nabla\nabla\tau(\bm\theta)$ are constrained by inequalities to enforce
  supplied image parities.  These are not known a priori, but have to be
  inferred by the investigator based on the brightness of the images, time-delay
  data, and test runs.
\item The time delay $\tau(\bm{\theta}_{i}) - \tau(\bm{\theta}_{j})$
  between images must reproduce the observed time delays.
\end{enumerate}

These equality and inequality constraints are all linear in the mass tiles
$\kappa_n$, the source position $\bm\beta$, the external shear
components $\gamma_1,\gamma_2$, any additional point masses, and
finally \Ho.  As a result, in the high-dimensional space of these
quantities, there is a convex polytope inside which all points satisfy
the prior and data constraints.  \cite{Lubini12} developed an
algorithm for uniformly sampling high-dimensional polytopes, which is
implemented in \GLASS.
The following definitions are used in GLASS modelling:
\begin{itemize}[leftmargin=*]
  \item A lensing mass $\Sigma$ is made up of $\sim$200 pixels, whose values are
  individually adjustable subject to the constraints discussed above.
  \item A \emph{model} is the set of a mass distributions $\Sigma_i$, shear
  terms, etc. for all eight lenses which reproduce the respective image and
  time-delay data and share a common value for $\Ho$.
  \item An \emph{ensemble} is a set of 1000 models drawn uniformly from the
  space of solutions that satisfy the above constraints.
\end{itemize}
A detailed discussion of the method and prior is given by \cite{Coles08},
and the consequences of the various assumptions have been studied
in previous works such as \cite{Lubini14,Kueng15,Denzel20}.

\subsection{Point-image data}

\tabref{td:systems} lists the point-image and time-delay data used in this
study.  We also indicate the locations of external galaxies approximated as
point masses.

  \begin{table*}
    \centering%

\begin{tabular}{lcrlllcl}
    \hline\hline
    System & Labels & $x, y$   & $z_L$ & $z_S$ & Time delays & Point mass      & Reference\\
           &        & [arcsec] &       &       &   [days]    & $x, y$ [arcsec] & \\
    \hline
    B1608+656 & B & (-0.82, \hphantom{-}1.28)            & 0.63 & 1.39 &                 & (\hphantom{-}0.14, \hphantom{-}0.79) & \tref{1}; \tref{2}; \tref{3}; \tref{4} \\
              & A & (\hphantom{-}1.10, \hphantom{-}0.42) &      &      & 31.5 $\pm$ 1.5, &   & \\
              & C & (\hphantom{-}0.68, \hphantom{-}1.20) &      &      &  4.5 $\pm$ 1.5, &   & \\
              & D & (-0.23, -0.64)                       &      &      & 41   $\pm$ 1.5  &   & \\
    \hline
    DESJ0408-5354 & A & (\hphantom{-}1.25, -2.15)             & 0.597 & 2.375 &                   &   & \tref{5}; \tref{6}\\
                  & B & (\hphantom{-}0.04, \hphantom{-}1.86)  &       &       & 112.1 $\pm$  2.1, &   & \\
                  & C & (\hphantom{-}1.26, \hphantom{-}1.77)  &       &       &  &   & \\
                  & D & (-1.37, \hphantom{-}0.15)             &       &       & 155.5 $\pm$ 12.8$^\dagger$,  &   & \\
                  &   &                                       &       &       & 42.4 $\pm$ 17.6$^\ddagger$   &   & \\
    \hline
    HE0435-1223 & C & (\hphantom{-}1.35, -0.36)             & 0.454 & 1.689 &                  &   & \tref{7}; \tref{8}; \tref{9}; \tref{10}\\
                & A & (-1.18, -0.07)                        &       &       & 2.1  $\pm$ 0.78, &   & \\
                & D & (-0.06, -1.09)                        &       &       & 6    $\pm$ 1.07, &   & \\
                & B & (\hphantom{-}0.19, \hphantom{-}1.13)  &       &       & 8.37 $\pm$ 1.31  &   & \\
    \hline
    PG1115+080 & C     & (\hphantom{-}0.11, \hphantom{-}1.37)  & 0.311 & 1.735 &                 &   & \tref{11}; \tref{12}; \tref{13}; \tref{14}\\
               & A$_1$ & (-0.79, -0.87)                        &       &       & 13.3 $\pm$ 0.9, &   & \\
               & A$_2$ & (-1.06, -0.43)                        &       &       &                 &   & \\
               & B     & (\hphantom{-}0.86, -0.46)             &       &       & 11.7 $\pm$ 1.5  &   & \\
    \hline
    RXJ0911+0551 & B     & (-2.27, \hphantom{-}0.29)             & 0.769 & 2.763 &              & (-0.76, \hphantom{-}0.66) & \tref{15}; \tref{16}; \tref{17}; \tref{18}\\
                 & A$_1$ & (\hphantom{-}0.96, -0.10)             &       &       & 146 $\pm$ 8, &   & \\
                 & A$_2$ & (\hphantom{-}0.70, -0.50)             &       &       &              &   & \\
                 & A$_3$ & (\hphantom{-}0.69, \hphantom{-}0.46)  &       &       &              &   & \\
    \hline
    RXJ1131-1231 & B & (\hphantom{-}1.39, \hphantom{-}1.64)  & 0.295 & 0.658 &                   & (\hphantom{-}0.51, -0.14) & \tref{7}; \tref{19}; \tref{20}; \tref{21}; \tref{22}; \tref{12}\\
                 & C & (-0.96, \hphantom{-}2.06)             &       &       &  1.50 $\pm$ 2.49, &   & \\
                 & A & (\hphantom{-}0.29, \hphantom{-}2.13)  &       &       &  9.61 $\pm$ 1.97, &   & \\
                 & D & (-0.23, -1.18)                        &       &       & 87    $\pm$ 8     &   & \\
    \hline
    SDSSJ1004+4112 & C & (\hphantom{-}8.85, -4.44)  & 0.68 & 1.74 &                  &   & \tref{23}; \tref{24}; \tref{25}; \tref{26}; \tref{27}\\
                   & B & (-5.76, -6.36)             &      &      & 821.6 $\pm$ 2.1, &   & \\
                   & A & (-2.46, -8.19)             &      &      &  40.6 $\pm$ 1.8, &   & \\
                   & D & (-2.37, \hphantom{-}4.64)  &      &      &                  &   & \\
    \hline
    WFIJ2033-4723 & B     &  (\hphantom{-}0.60, \hphantom{-}1.35)  & 0.661 & 1.66 &                 &   & \tref{28}; \tref{29}; \tref{30}\\
                  & A$_1$ &  (\hphantom{-}0.31, -1.21)             &       &      & 35.5 $\pm$ 1.4, &   & \\
                  & A$_2$ &  (\hphantom{-}0.86, -0.66)             &       &      &                 &   & \\
                  & C     & (-0.88, -0.23)                         &       &      & 27.1 $\pm$ 1.4  &   & \\
    \hline
  \end{tabular}
    \caption{\tablbl{td:systems}%
    Input parameters used to model the gravitational lens systems.  
    The images of each lens are ordered according to arrival times (earliest on
    top).  Time delays have been recalculated accordingly. If not otherwise
    indicated, all time delays are with respect to the previous image.  Image
    coordinates $x, y$ are relative to the center of the lensing galaxy.  The
    measured redshifts $z_L$ and $z_S$ correspond to the lens and source
    redshifts.  To account for any significant external lensing contribution
    (besides external shear), we added point masses at the positions where
    galaxies of similar redshift are visible in the field. Every lens has been
    subjected to rigorous investigations in the past as explained in
    \secref{td:systems}; the references to the works used particularly in this
    study are cited here:
    \protect\tlink{1}  \protect\cite{B1608Koopmans99}; 
    \protect\tlink{2}  \protect\cite{B1608Fassnacht99};
    \protect\tlink{3}  \protect\cite{B1608Fassnacht02};
    \protect\tlink{4}  \protect\cite{B1608Koopmans03};
    \protect\tlink{5}  \protect\cite{DESJ0408Lin17};
    \protect\tlink{6}  \protect\cite{DESJ0408Courbin18};
    \protect\tlink{7}  \protect\cite{Millon20a};
    \protect\tlink{8}  \protect\cite{HE0435Bonvin17};
    \protect\tlink{9}  \protect\cite{HE0435Sluse12};
    \protect\tlink{10} \protect\cite{HE0435Kochanek06};
    \protect\tlink{11} \protect\cite{PG1115Morgan08};
    \protect\tlink{12} \protect\cite{PG1115Tonry98};
    \protect\tlink{13} \protect\cite{PG1115Barkana97};
    \protect\tlink{14} \protect\cite{PG1115Weymann80};
    \protect\tlink{15} \protect\cite{Eulaers11};
    \protect\tlink{16} \protect\cite{RXJ0911Hjorth02};
    \protect\tlink{17} \protect\cite{RXJ0911Burud98};
    \protect\tlink{18} \protect\cite{RXJ0911Bade97};
    \protect\tlink{19} \protect\cite{RXJ1131Birrer16};
    \protect\tlink{20} \protect\cite{RXJ1131Tewes13};
    \protect\tlink{21} \protect\cite{RXJ1131Sluse07};
    \protect\tlink{22} \protect\cite{RXJ1131Sluse03};
    \protect\tlink{23} \protect\cite{SDSSJ1004Fohlmeister07};
    \protect\tlink{24} \protect\cite{SDSSJ1004Fohlmeister08};
    \protect\tlink{25} \protect\cite{SDSSJ1004Williams04};
    \protect\tlink{26} \protect\cite{SDSSJ1004Oguri04};
    \protect\tlink{27} \protect\cite{SDSSJ1004Inada03};
    \protect\tlink{28} \protect\cite{WFI2033Bonvin19};
    \protect\tlink{29} \protect\cite{WFI2033Eigenbrod06};
    \protect\tlink{30} \protect\cite{WFI2033Morgan04}. \\ 
    $\dagger$: With respect to the first image.\hspace{1em}
    $\ddagger$: With respect to the second image.  }%
  \end{table*}%

The following settings were also used.
\begin{itemize}[leftmargin=*]
  \item The mass pixels were arranged in a discretised circular disc 17 pixels
  in diameter.  The central pixel was further refined into $3\times3$
  sub-pixels.  The side length of the mass pixels was between $0.19''$ and
  $0.35''$ with SDSSJ1004+4112 being an exception at $1.4147''$.  Note that
  $\tau(\bm\theta)$ is not discretised.
  \item All models allowed for external shear.
  \item A flat cosmology with $(\Omega_{\text{m}}, \Omega_{\Lambda}) = (0.27,
  0.73)$ was assumed.
\end{itemize}

\subsection{Extended image data}\subseclbl{td:extended_images}

The procedure thus far uses solely the centroidal positions of the multiply
imaged quasars.  We now incorporate the full photometric data of the extended
images as described in \cite{Denzel20} and related earlier work \citep{Kueng18}.

As images appear where $\nabla\tau(\bm\theta)=0$, we first define a lens
mapping $L(\bm\theta,\bm\beta)$, which is equivalent to the more commonly
known lens equation.  An extended image produced by an extended source can then
be written as
\begin{equation}\eqlbl{td:extimage}
  I(\bm\theta) = \int L(\bm\theta,\bm\beta) \, S(\bm\beta) \, d^2\bm\beta \,.
\end{equation}
For each of the $1000\times8$ lens masses in an ensemble we generate synthetic
images by fitting a source brightness distribution $S(\bm\beta)$ such that the
extended image $I$ further convolved by a point spread function
$P(|\bm\theta-\bm\theta'|)$ 
\begin{equation}
  \int P(|\bm\theta-\bm\theta'|) \, L(\bm\theta',\bm\beta) \,
       S(\bm\beta) \, d^2\bm\theta' \, d^2\bm\beta
\end{equation}
optimally fits the data.  
The extended image is linear in the source brightness distribution, so fitting
is straightforward.  

\subsection{Lessons from the TDLMC}\subseclbl{td:tdlmc}

In January 2018, the Time Delay Lens Modelling Challenge was
initiated.  \cite{TDLMC1} explained the experimental design and
invited scientists to model 50 simulated Hubble Space Telescope (HST)
observations of mock lens systems.  The challenge was divided into
three `rungs' each featuring a set of 16 lenses which increased in
modelling difficulty.  Additionally, the `Evil' team, the designers of
the challenge, provided a calibration rung containing 2 lenses along
with the entire information about the systems.  For all other lenses,
an HST mock image of the lens, the redshifts of source and lens,
external convergence estimates, velocity dispersion estimates, the
time delays, a noise map, and a point-spread function (PSF) map were
provided.  Based on this information, the challenge for the `Good'
teams was to infer \Ho{}, which was randomly fixed between 50 and
90~\Hunits{} by the Evil team, for each rung for a fixed background
cosmology.  With final submission on September 8th 2019, the TDLMC
finished its submission phase, and the results were thoroughly
evaluated \citep{TDLMC2}.

We have participated in this challenge and were able to test our
free-form modelling techniques extensively with focus on \Ho{}
inference beforehand.  Lessons learned from the challenge influenced
the present work in several ways.

Tests on simulated data always come with some caveats.  In particular,
mock lenses in rung3 were derived from hydrodynamical simulations of
galaxies.  Expectedly, this limits the quality of mock lenses to the
one of the used simulations.  In particular, the resolution of the
galaxies were known to be insufficient to resolve the inner regions of
the galaxies.  Some of the limitations on the other hand were
discovered post-unblinding, such as inconsistencies with the
kinematics due to the removal of substructure, and the halo truncation
at the virial radius, which does not follow isodensity contours and
potentially biases \Ho{}.

More importantly, during the challenge we have noticed that the
simulated lenses differ from observed ones in at least one important
aspect: the radial spread of the images around the lens center is
considerably wider, by about a factor of 2, in observed lenses
compared to the simulated ones (also see Gomer~\&~Williams~2020,
in preparation). This leads us to expect that the lens sample from
this work should yield more precise results on the inference of \Ho{}.
In the last paragraph of \subsecref{td:H0} and \figref{td:tdlmc} we
present the accuracy and precision of our TDLMC results, and compare
the precision to that of the present work.

Our TDLMC participation was also helpful in guiding the modelling
procedure carried out in this paper.  The most striking conclusion we
drew from our TDLMC participation was that as soon as double systems
entered the analysis, our \Ho{} inference tended to much lower and
less accurate values; at times even lower than the asserted range of
possible values.  After rung1 however, we decided to include a
joint-analysis containing only quad systems, besides other
combinations of quads and doubles.  The best results were achieved
this way and are shown as histograms of our \Ho{} posteriors in
\figref{td:tdlmc}. This was the main reason why only quad systems were
selected for the sample presented in this work.

\section{Results}\seclbl{td:results}

Most of the results in this paper come from an ensemble of 1000 eight-lens
models required to reproduce the point-image and time-delay data.  This ensemble
was also subsequently filtered according to how well extended images could be
reproduced.  We consider the average of the ensemble, and also the variation
within the ensemble.  Because the model ensemble is constructed using linear
constraints, any weighted average of ensemble members is also a valid model.

Apart from the main ensemble, we produced two further models using the same
image and time-delay data, but constrained to the \Ho{} values from Planck and
SH0ES mentioned at the beginning of this paper, to demonstrate their
compatibility.  Furthermore, we compare distribution of inferred \Ho{} to models
for the mock data from the TDLMC in \subsecref{td:H0}.

\subsection{Arrival-time surfaces}\subseclbl{td:arrivs}
\figref{td:arriv} shows contour maps of the arrival time $\tau(\bm\theta)$ from
the average models of the main ensemble.  The orientation in this figure is
different from \protect\figref{td:composites} but consistent with subsequent
figures.  In the classification of \cite{Saha03} HE0435-1223 is a core quad,
RXJ1131-1231 a long-axis quad, RXJ0911+0551 a short-axis quad, while the rest
are inclined quads.  The minima and saddle point appear at the correct image
locations, as required, and there are no indications of spurious additional
images.
  
\subsection{Convergence maps}\subseclbl{td:kappas}
\figref{td:kappa} shows convergence maps $\kappa(\bm\theta)$ of the
ensemble-average models.  The maps exhibit the typical pixellated structure
stemming from the free-form technique.  Contrary to earlier work imposing
inversion symmetry in most cases \citep{Saha06b,Paraficz10} all the lens models
allow for asymmetry, which seems to be an important feature, in e.g.,
DESJ0408-5354 and RXJ0911+0551. Rough shapes and orientations seem to agree with
previous reconstructions \citep{Chantry10, Wynne18, Shajib20, PG1115Yoo05,
Saha07}.  The area within the black contours on the maps indicate a
supercritical density with $\kappa \geq 1$.  This area defines a scale which can
be expressed as the Einstein radius and formally corresponds to the radius where
$\langle\kappa\rangle_{\mathrm{R}_{\text{E}}} = 1$.
  
In \figref{td:profiles} the value for the Einstein radius can easily be read
off (as vertical lines). It shows the distribution of average enclosed
$\kappa$ as a function of radial distance from lens centre for the entire
ensemble. The ensemble-average profile is depicted in red.  As it is typical
for (good) lens reconstructions, the spread in the Einstein radius is minimal
and usually accurate due to the strong constraints coming from the images
themselves which usually lie at comparable radii. Conversely, the spread
within the ensemble towards center is much greater ranging from cored profiles
to almost cusp-like centres. This too was expected due to the lack of
constraints at these radii.  Note that all profiles are in units of $\kappa$
which is the surface density scaled to the critical density for each lens,
giving them the appearance of having seemingly similar mass contents.
Particularly for SDSSJ1004+4112, the masses exceed magnitudes of (normal)
galaxies and clearly reach cluster scales.

Since free-form lens-reconstruction ensembles can in principle contain many
differently shaped density maps, it is important to investigate dominant as well
as secondary features across all models in an ensemble.  As explained in
\secref{td:methods}, these models represent solutions from a high-dimensional
space, treating each mass tile as a parameter.  To do this we carried out a
principal-components analysis (PCA) of the ensemble \citep[similar
to][]{SDSSJ1004Mohammed15,Kueng18}.  A PCA yields a representation of the mass
distribution for the $k$-th lens ($k=1,2,\ldots,8$) of the form
\begin{equation}
  \kappa^{(k)}(\bm\theta) = \bar\kappa^{(k)}(\bm\theta) +
  {\textstyle\sum_p} c_p \, \Delta\kappa^{(k)}_p(\bm\theta)
\end{equation}
where $\bar\kappa^{(k)}$ is the ensemble-average lensing mass (as shown in
\figref{td:kappa}), while $\Delta\kappa^{(k)}_p$ is the $p$-th principal
component, and $c_p$ is a coefficient.  The $\Delta\kappa^{(k)}_p$ are
orthonormal by construction.  Note that the coefficients $c_p$ do not depend on
$k$, but are common for all eight lenses.  Each of the $c_p$ has a range of
values across the ensemble: the coefficient $c_1$ of the first principal
component has the largest range, while $c_{100}$ stays close to zero.  Thus, PCA
filters and sorts the ensemble for its most salient features.  \figref{td:pca}
shows an example of lens models projected into the PCA-feature space.  It
considers the lens WFIJ2033-4723 and the 1st, 2nd, 5th, and 100th principal
components.  For each of these, we display the projection
\begin{equation}\eqlbl{td:pcaproj}
\bar\kappa^{(k)}(\bm\theta) +
c_p \, \Delta\kappa^{(k)}_p(\bm\theta) \,.
\end{equation}
for the 16th and 84th percentile values of $c_p$.  This figure illustrates the
variety of models within an ensemble, but concurrently also identifies regions
in the convergence maps which have been constrained to a higher degree and thus
have low variance.  As observed in \figref{td:profiles}, the convergence maps
are typically well constrained around the notional Einstein ring (indicated by a
black contour in Figures~\figref*{td:kappa} and \figref*{td:pca}), since in most
cases by the 5th principal component its shape already does not vary anymore.

\subsection{Synthetic images}\subseclbl{td:synthetics}
Another rather affirming result is presented by the synthetic images from source
reconstructions using the ensemble-averaged lens models.  In the least-squares
fitting discussed in \subsecref{td:extended_images}, a Poisson noise in the
photometry was assumed $\sigma_{\bm{\theta}}^{2} = g^{-1}
|D^{\text{obs}}_{\bm{\theta}}|$ where $g$ is the gain or counts per photon.
The source reconstructions yielded reduced $\chi^{2}$ which are listed in
\tabref{td:chi2}.

  \begin{table}
    \begin{tabular}{lcrrr}
      \hline\hline
      System & Date & $\chi^{2}$ & $\chi^{2}_{\text{Planck}}$ & $\chi^{2}_{\text{SH0ES}}$ \\
      \hline
      B1608+656      & 2004-08-24 &   1.24 &   1.22 &   1.26 \\
      DESJ0408-5354  & 2018-01-17 &   3.90 &   4.01 &   3.78 \\
      HE0435-1223    & 2011-04-11 &   3.88 &   3.70 &   3.88 \\
      PG1115+080     & 2013-03-23 &   1.99 &   1.90 &   1.99 \\
      RXJ0911+0551   & 2012-10-19 &   8.97 &   9.78 &   9.17 \\
      RXJ1131-1231   & 2004-06-24 &   8.18 &   9.63 &   8.94 \\
      SDSSJ1004+4112 & 2010-06-07 & 102.67 & 102.53 & 100.74 \\
      WFIJ2033-4723  & 2013-05-03 &   6.17 &   6.27 &   5.93 \\
      \hline
    \end{tabular}
    \caption{Reduced $\chi^{2}$ from the source-reconstruction fitting. The
      first column refers to least squares from the main ensemble model's
      average. $\chi^{2}_{\text{Planck}}$ and $\chi^{2}_{\text{SH0ES}}$ are the
      corresponding least squares from the ensembles which have been constrained
      to fixed values of \Ho{} from Planck and SH0ES.}
    \tablbl{td:chi2}
  \end{table}

  The reduced least-squares provide a standardized measure of the synthetic's
  quality. A $\chi^{2}=1$ means the synthetic image differs from the data only
  on noise level and provides an ideal fit. However, since the observations come
  from different camera systems, wavelength bands, and dates ranging from 2004
  to 2018, the data contains various signal-to-noise ratios (SNR). Consequently,
  a $\chi^{2}$ of say 1.5, is easier to reach for a relatively old and noisy
  image, compared to one taken with a more modern system with a lower SNR.  The
  scatter of $\chi^{2}$ within a lens system is typically only between 1 or 2,
  in the case of SDSSJ1004+4112 about 8.  The least squares from the source
  reconstructions of the models which were constrained with values of \Ho{} from
  Planck and SH0ES (see \secref{td:intro}) were labelled
  $\chi^{2}_{\text{Planck}}$ and $\chi^{2}_{\text{SH0ES}}$.

  The synthetic images from averages of the main ensemble for each lens are
  shown in \figref{td:synthA}. Apart from a few minor details, all lensed
  features have successfully been reconstructed with astonishingly low
  $\chi^{2}$ (SDSSJ1004+4112 being an exception). In some cases, e.g. for
  B1608+656, RXJ0911+0551, and WFIJ2033-4723, the shape of the quasar images are
  slightly warped, perhaps due to too high shear components.  In B1608+656 and
  WFIJ2033-4723, flux ratios between some of the quasar images also noticeably
  differ. RXJ1131+01231, although the quasar images and most of the fainter
  Einstein ring are fitted relatively well for this system, shows artifacts
  which contribute the majority of errors. The origin of the artifacts are
  unclear.
  
  The synthetic image of SDSSJ1004+4112 shows much fainter images than the
  original.  This probably comes from the fact that the cluster galaxies induce
  much higher deflection angles and therefore more space for errors than for
  much smaller strongly-lensing galaxy systems.  Problems with SDSSJ1004+4112
  were expected since it is a cluster lens with many details which have been
  missed during the modelling.  The considerably higher $\chi^{2}$ in
  \tabref{td:chi2} compared to the other systems are a sign of such problems.
  In part, the high $\chi^{2}$-values are explained by bad fits of the lower
  brightness regions around the notional Einstein radius.  These differences are
  discernible in \figref{td:synthB} upon close inspection of the original and
  the synthetic image side by side. Another source of errors are the cluster
  galaxies which have not been masked properly.

  \subsection{\Ho{} posterior distribution}\subseclbl{td:H0}
  In \figref{td:H0} and \figref*{td:invH0}, the posterior distribution of values
  for \Ho{} and $\Ho^{-1}$ from the main ensemble model are depicted. These do
  not yet include constraints from the extended image data.  We furthermore
  investigated whether the quality of source reconstructions correlated with
  \Ho{}.  For this reason, we optimized the ensemble models in a post-processing
  step by calculating synthetic images for each model in the ensemble (as
  described in the previous subsection). Only a fraction of the models with the
  best overall $\chi^{2}$ (excluding SDSSJ1004+4112) was kept and the rest
  discarded. Several fractions were tested, ranging from 10 to 80 per-cent.  The
  \Ho{} posterior of the 30 per-cent-filtered ensemble is shown in
  \figref{td:H0filtered}.  Interestingly, this had no noticeable effect on the
  spread, no matter how many models were filtered out. This suggests that better
  source reconstructions of time-delay lenses will not place tighter constraints
  on \Ho{}.  In particular, \figref{td:H0filtered} shows the median of the
  distributions at
  $$\Ho = \HaHzres{} = \Hres{}$$
  and 
  $$\Ho^{-1} = \invHres,$$
  each at 68\% confidence.
  
  The Hubble constant is also equivalent to the cosmological critical
  density density $\rho_c$ (see \ref{eq:td:rho_crit}).  The
  distribution of $\rho_c$ values in the main ensemble is displayed in
  \figref{td:rhoH0} in units of $\mathrm{GeV}/\mathrm{m}^3$.  Its
  median has a value of $\rho_c = \rhocritres$, at 68\%
  confidence. For values of \Ho{} ranging from 60 to 80~\Hunits\, the
  critical density corresponds to roughly 1 to 2 alpha particles per
  cubic metre.  This is the quantity which should be compared to the
  early measurements, as those generally constrain \Ho{} through the
  baryon or matter densities $\propto\Ho{}^2$.

  The spread in the distribution of \Ho{} is large, in fact the ensemble contains
  solutions ranging from 60 to more than 80~\Hunits{}. This means that any value
  of \Ho{} is in principle consistent with the data constraints and priors. The
  reason for this lies for the most part in how the free-form technique
  builds convergence maps.  It explores the degenerate solution space for the
  lens equation coupled with a few physical and regularization priors.  At
  first, this spread could imply that the solutions have not been properly
  constrained or regularized.  However, as shown in the previous
  \subsecref{td:synthetics}, the models are able to reproduce the extended lens
  photometry quite well (with only one exception). This suggests that the models
  are in fact on average physically viable. Note that the extended lens
  photometry didn't go into the modelling process.

  Another interesting observation about the \Ho{} distribution is that its error
  distribution does not appear to be Gaussian, in fact the distribution in
  \figref{td:H0filtered} is clearly asymmetric.  In astronomy, most analyses
  generally use a Gaussian error distribution, firstly, because of the central
  limit theorem, and secondly, because the assumption simplifies the estimation
  of unknown parameters.  However, with realistic data, we do not know the
  probability distribution of the errors, nor whether it has any concrete
  mathematical form consistent from one observation to another.  The wings of a
  Gaussian fall off quickly, meaning two or three $\sigma$ residuals are very
  unlikely to occur.  From experience, however, we know such deviations are far
  more common. Thus, an error distribution such as a Lorentzian or a Voigt
  profile which have a well-defined peak with wider wings, might be more
  reasonable estimates for realistic data.  Even more flexible is the Tukey
  g-and-h \citep{tukeygh} distribution which allows for asymmetric wings.
  Gaussian distributions arise naturally when a quantity is a sum of
  many independent contributions and physical processes.  Thus noise
  is typically Gaussian, and so are measurements that are signals on
  top of noise.  The inference of the Hubble constant from
  observations of several lenses cannot, however, be decomposed into
  signal plus noise.  Hence the uncertainty in \Ho{} is not in general
  Gaussian, and one can expect non-Gaussian properties such as
  skewness and kurtosis to be important.  Likewise, a 1 per-cent
  determination of the Hubble constant which implicitly uses Gaussian
  errors, might actually have overestimated the precision of the
  measurement.

  Comparing the spread of the distribution of \Ho{} values in the present work,
  \figref{td:H0filtered} and \figref*{td:H0}, to the ones from the TDLMC, it is
  apparent that this it is considerably narrower.  Part of the reason for this
  could be the difference in the radial spread of images around the lens center,
  mentioned in \subsecref{td:tdlmc}. If images are confined to a narrow band
  around the lens center, as in TDLMC quads, lensing degeneracies, like the
  steepness, or mass sheet degeneracy, will be more rampant. This is because a
  given observational uncertainty on time delays will lead to larger fractional
  uncertainty in more circular lenses which have smaller time delays, resulting
  in less constraining power, and a wider range of derived \Ho{} values.
  Observed quads span a wider range of radial image positions, resulting in
  narrower \Ho{} distributions.
 
  While our TDLMC models have a very large spread across the entire range of
  possible \Ho{} values, their median \Ho{} seems to determine the truth values
  quite accurately.  This does not mean that our techniques are able to
  consistently recover the truth values, but it still shows potential.

\section{Conclusion}\seclbl{td:conclusion}

The Hubble constant has come a long way from the value of $\sim500\,\Hunits$
implied by the historic Figure~1 in \cite{Hubble29}.  \cite{Sandage58} had
improved the measurement to $\Ho\approx75\,\Hunits$ or
$\Ho^{-1}\approx\rm\,13\,\Gyrs$ with an uncertainty of a factor of two.  Debates
over a factor of two were still continuing in \cite{oldH0low} versus
\cite{oldH0high}.  Today, \Ho\ has been constrained to within 10 per-cent, but
the debates between ``low'' and ``high'' values remain, as do the legacy units
of \Hunits.  In view of the continuing debate over the value of \Ho, lensing
time delays as a technique for measuring \Ho\ are very interesting, because they
span a variety of redshifts and are free of the ``twilight zone''
\citep[see][]{twilightzone} characteristic of the distance-ladder methods.

In this paper, we have presented a determination of the Hubble
constant $\Ho{} = \HaHzres{} = \Hres{}$ through the joint free-form
modelling of eight time-delay lenses using the most recent
observational data.  This value is consistent with both early and late
Universe studies. We further demonstrate this fact by modelling these
lenses in secondary ensembles which have fixed values of \Ho{} typical
for CMB-based and SNeIa-based methods.  Accordingly, these secondary
ensemble models exhibit just as many or few problems as the main lens
ensemble.  We have analysed the models based on their arrival-time
surfaces, convergence maps, and circularly-averaged density profiles
and have found only minor shortcomings.  Furthermore, the models'
projection properties have been tested by generating synthetic images
using the source-reconstruction method presented in \cite{Denzel20}.
Thereby, we have extended our analysis to data which have not
initially been employed in the lens reconstructions. The results of
this test affirmed the physical validity of our models, and so, the
synthetic images and thereby the entire photometric data are finally
used to constrain the ensemble further in a post-processing step. This
had no notable effect, suggesting that optimizations of source
reconstructions only weakly constrain \Ho{}.  In our study, we have
not (yet) considered phase-space models, or stellar kinematics, which
might be able to further constrain our models.

As of the time of writing, a 1\% determination of \Ho{} through
lensing has yet to be reported.  Nevertheless, even if such a
measurement existed, only several repetitions with different data sets
could confirm the accuracy of these measurements.  As already
discussed in \subsecref{td:H0}, when we assess data, it is entirely
unknown what kind of error distribution can be assumed.  Thus, only if
several measurements of \Ho{} across different observations
consistently reach a 1 per-cent level, they are robust against the
choice of errors.  However, the TDLMC \citep{TDLMC2} suggests that
while many lens-modelling techniques excel in reconstructing simple
simulations based on parametric models, but decrease in consistency
when faced with more general lenses taken from galaxy-formation
simulations. As cautioned in several works
\citep[e.g.][]{Gomer19,Kochanek20} due to lensing degeneracies a
single family of models is able to reproduce the same lensing
observables, but return different values of \Ho{}. Free-form
techniques keep consistency in accuracy and precision when the
complexity of the lens is increased, as expected from their greater
flexibility.

Interestingly, our new estimate of \Ho{} improves upon the precision
of our measurements reported in the TDLMC. In the challenge, the
simulated quads appeared to be slightly rounder with very little
variation in the radial distance of the images.  The real observations
considered in this study appear to have a larger variation.  We
conjecture that a wider radial distribution of the lensed images puts
tighter constraints on the slope of the density profile, and therefore
provides tighter constraints on the Hubble constant.  From our
experience in the TDLMC, we also conclude that the most accurate joint
inference of \Ho{} comes from modelling only quad systems.  In
addition to the more elliptical lens-image separations, we have
improved upon the TDLMC in precision by increasing the number of
simultaneously modelled quad systems from 4 to 8.  This implies that
minor improvements might be possible by increasing the number of
modelled systems alone, provided they contain new information which is
able to further constrain \Ho{}.

It may turn out that in galaxy lenses degeneracies impose a limit
which can hardly be broken, and results similar to \Ho{} reported in
this study perhaps are the best we can hope to attain.  If this turns
out to be the case, time delays for galaxy lenses can still be useful,
by reversing the problem and using them together with \Ho{} from other
methods to constrain substructure in
lenses~\citep{SDSSJ1004Mohammed15}.

The situation is different in cluster lenses. Recently, \cite{Ghosh20}
considered cluster lenses with a time-delay quasar (such as
SDSS~J1004+4112 in this work) but also hundreds of other lensed
sources at many redshifts, making up $\sim1000$ images in all.  With
that many lensing constraints (expected to be achieved by JWST), a 1\%
measurement of \Ho{} from lensing time delays appears feasible.

%

%
\section*{Acknowledgments}
We would like to thank the anonymous referee for useful suggestions which
improved the paper.  PD acknowledges support from the Swiss National Science
Foundation.  This research is based on observations made with the NASA/ESA
Hubble Space Telescope obtained from the Space Telescope Science Institute
(STScI), which is operated by the Association of Universities for Research in
Astronomy, Inc., under NASA contract NAS 5–26555. These observations are
associated with programs \#10158, \#15320, \#12324, \#12874, \#9744, and
\#10509.

\section*{Data Availability}
The data underlying this article are available at the STScI
(\href{https://mast.stsci.edu/}{https://mast.stsci.edu/}; the unique
identifiers are cited in the acknowledgements). The derived data
generated in this research will be shared on request to the
corresponding author.

%
\bibliographystyle{mnras}
\bibliography{refs}

%

\newcommand{\mosaicgassym}[2]{
  \centering
  \includegraphics[#2]{\home/imgs/B1608+656_#1}\includegraphics[#2]{\home/imgs/DESJ0408-5354_#1}\includegraphics[#2]{\home/imgs/HE0435-1223_#1}\\%
  \includegraphics[#2]{\home/imgs/PG1115+080_#1}\includegraphics[#2]{\home/imgs/RXJ0911+0551_#1}\includegraphics[#2]{\home/imgs/RXJ1131-1231_#1}\\%
  \includegraphics[#2]{\home/imgs/SDSSJ1004+4112_#1}\includegraphics[#2]{\home/imgs/WFIJ2033-4723_#1}\\%
}

\newcommand{\mosaicgr}[2]{
  \centering
  \includegraphics[#2]{\home/imgs/B1608+656_#1}\includegraphics[#2]{\home/imgs/DESJ0408-5354_#1}\\%
  \includegraphics[#2]{\home/imgs/HE0435-1223_#1}\includegraphics[#2]{\home/imgs/PG1115+080_#1}\\%
  \includegraphics[#2]{\home/imgs/RXJ0911+0551_#1}\includegraphics[#2]{\home/imgs/RXJ1131-1231_#1}\\%
  \includegraphics[#2]{\home/imgs/SDSSJ1004+4112_#1}\includegraphics[#2]{\home/imgs/WFIJ2033-4723_#1}\\%
}

\newcommand{\pcamosaic}[2]{
  \centering
  \includegraphics[#2]{\home/imgs/#1_kappa_pca1_16}\includegraphics[#2]{\home/imgs/#1_kappa_pca1_84}\\%
  \includegraphics[#2]{\home/imgs/#1_kappa_pca2_16}\includegraphics[#2]{\home/imgs/#1_kappa_pca2_84}\\%
  \includegraphics[#2]{\home/imgs/#1_kappa_pca5_16}\includegraphics[#2]{\home/imgs/#1_kappa_pca5_84}\\%
  \includegraphics[#2]{\home/imgs/#1_kappa_pca100_16}\includegraphics[#2]{\home/imgs/#1_kappa_pca100_84}\\%
}

\newcommand{\synthpair}[2]{
  \includegraphics[#2]{\home/imgs/#1_data}\includegraphics[#2]{\home/imgs/#1_synth}%
}
\newcommand{\synthmosaicA}[1]{
  \synthpair{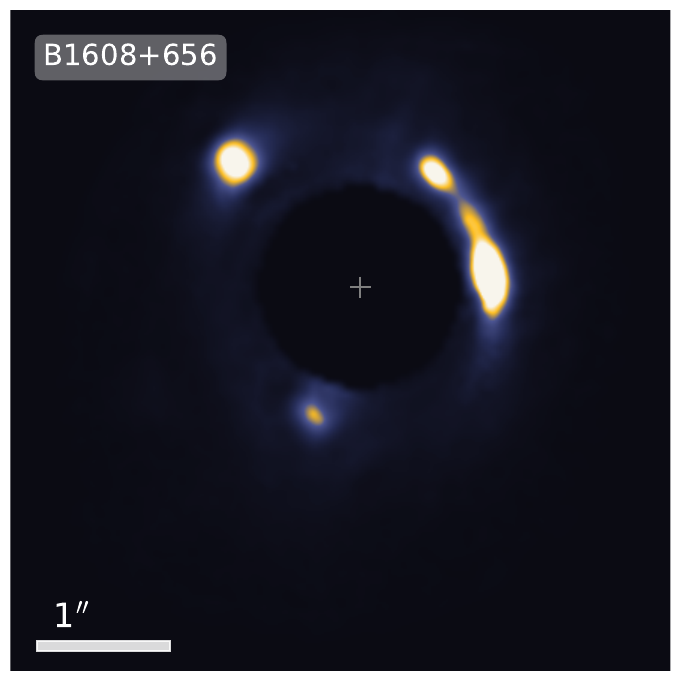}{#1}\\%
  \synthpair{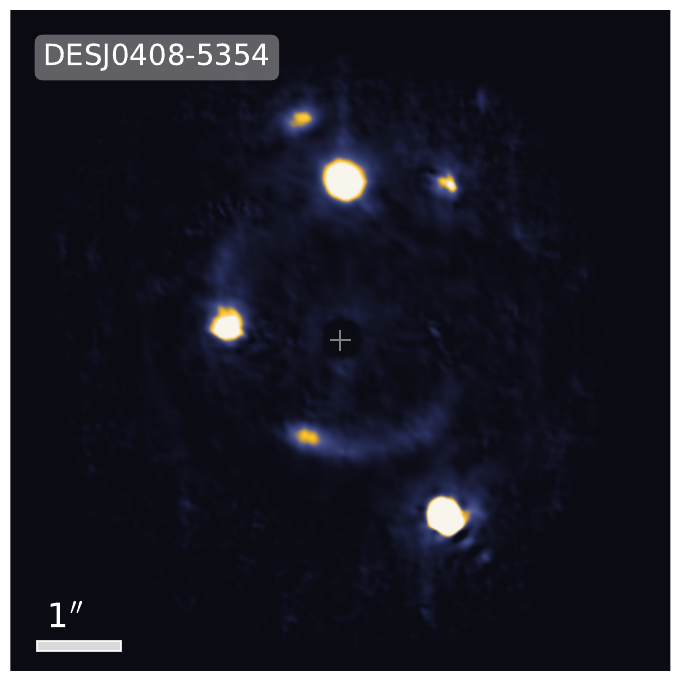}{#1}\\%
  \synthpair{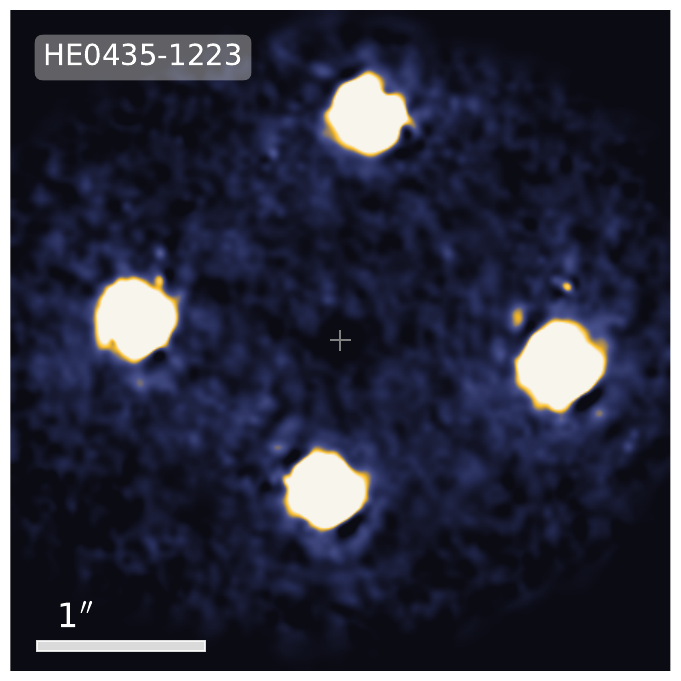}{#1}\\%
  \synthpair{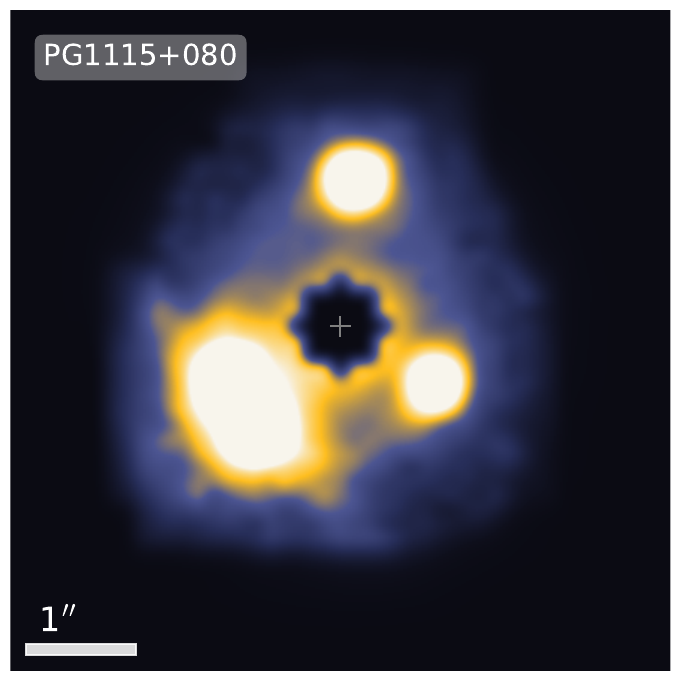}{#1}\\%
}
\newcommand{\synthmosaicB}[1]{
  \synthpair{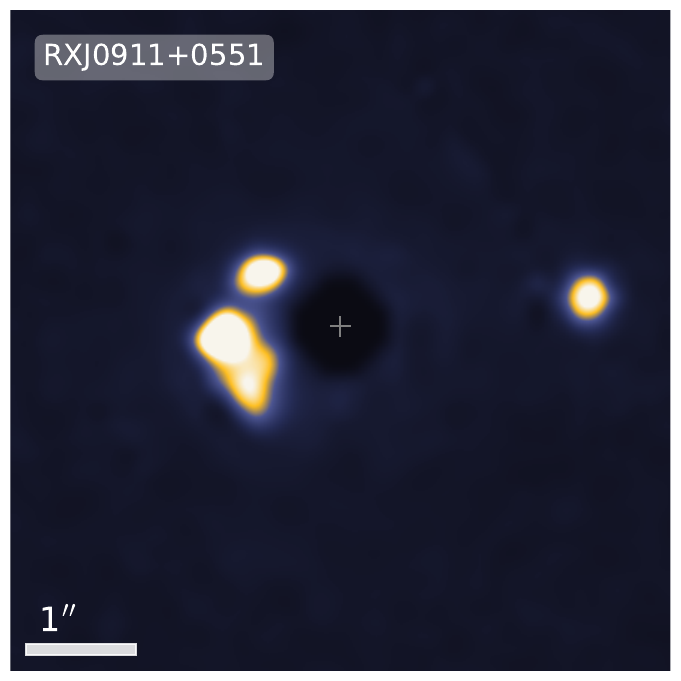}{#1}\\%
  \synthpair{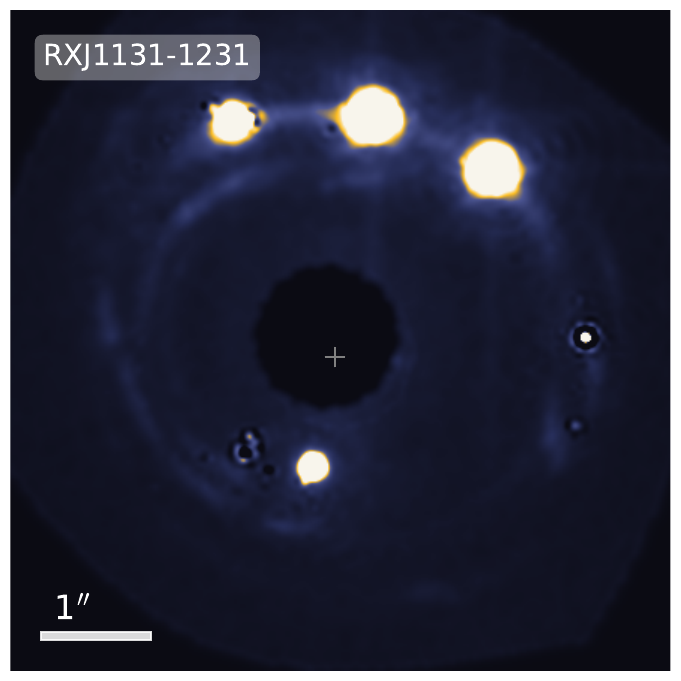}{#1}\\%
  \synthpair{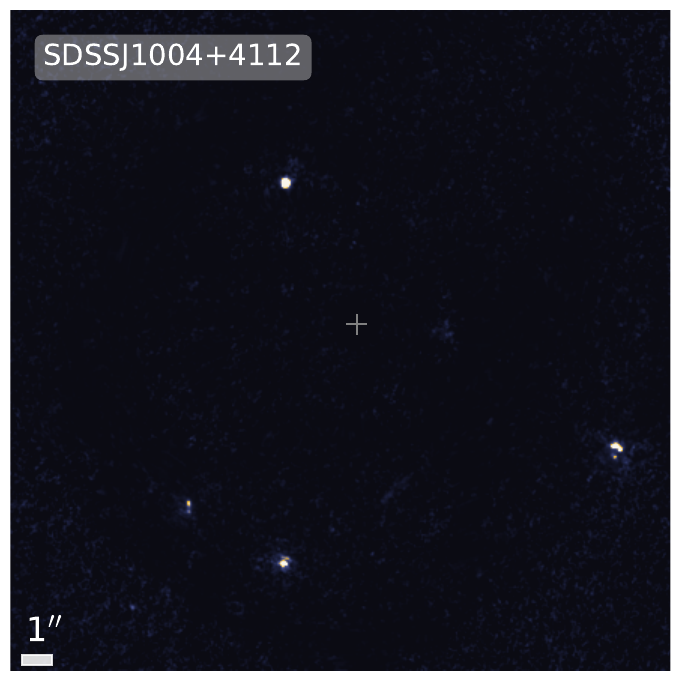}{#1}\\%
  \synthpair{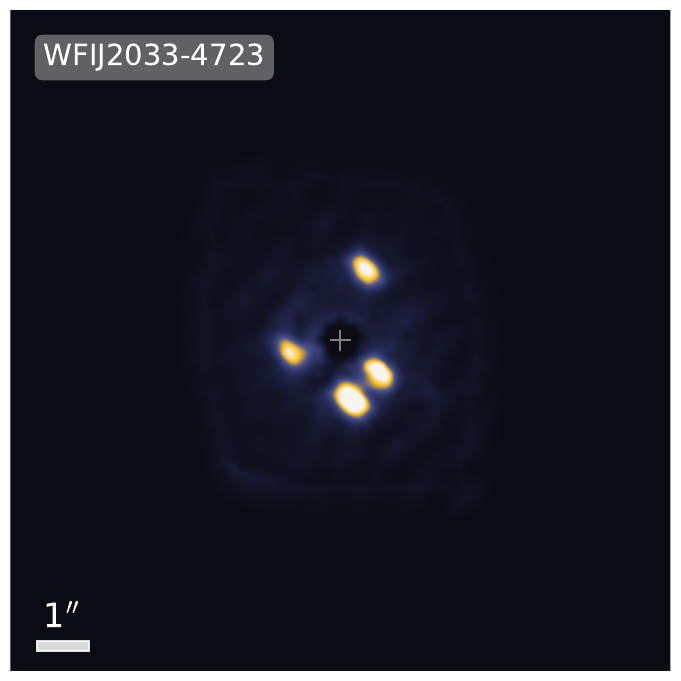}{#1}\\%
}

\section{Figures}\seclbl{td:figures}

  This section contains all figures which are referenced in previous sections.

  \begin{figure*}
    \mosaicgr{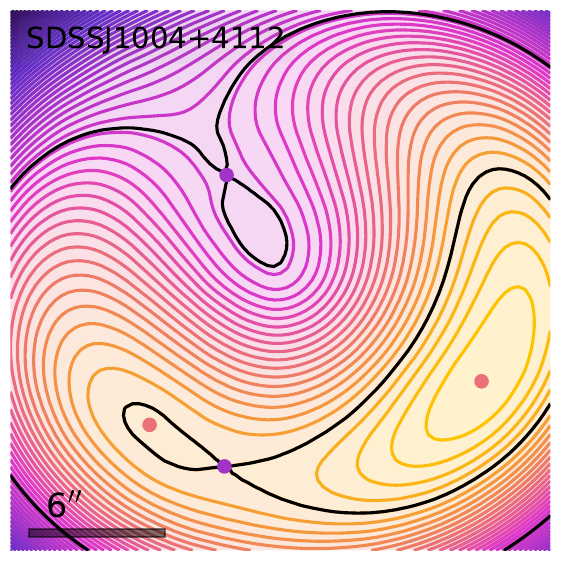}{height=\qheight}
    \caption{Arrival-time surfaces of the ensemble-average models.
      The image-position constraints with minimum and saddle parity
      are indicated by red and purple dots respectively.  Contours
      passing through saddle points are in black.  The scale bar on
      the lower left in each panel shows the angular scale in
      arcseconds.  Orientations are at arbitrary rotation compared to
      \protect\figref{td:composites}, but consistent with the following
      figures.   }\figlbl{td:arriv}
\end{figure*}

  \begin{figure*}
    \mosaicgr{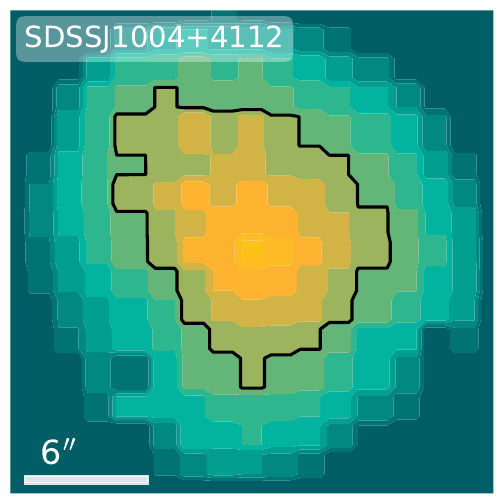}{height=\qheight}
    \caption{Model-convergence maps (ensemble averages) of all lenses.  Black
      contours indicate a $\kappa = 1$.  Scales and orientations are identical
      to the corresponding panels in \protect\figref{td:arriv}.
      }\figlbl{td:kappa}
\end{figure*}

  \begin{figure*}
    \includegraphics[height=0.9\textheight]{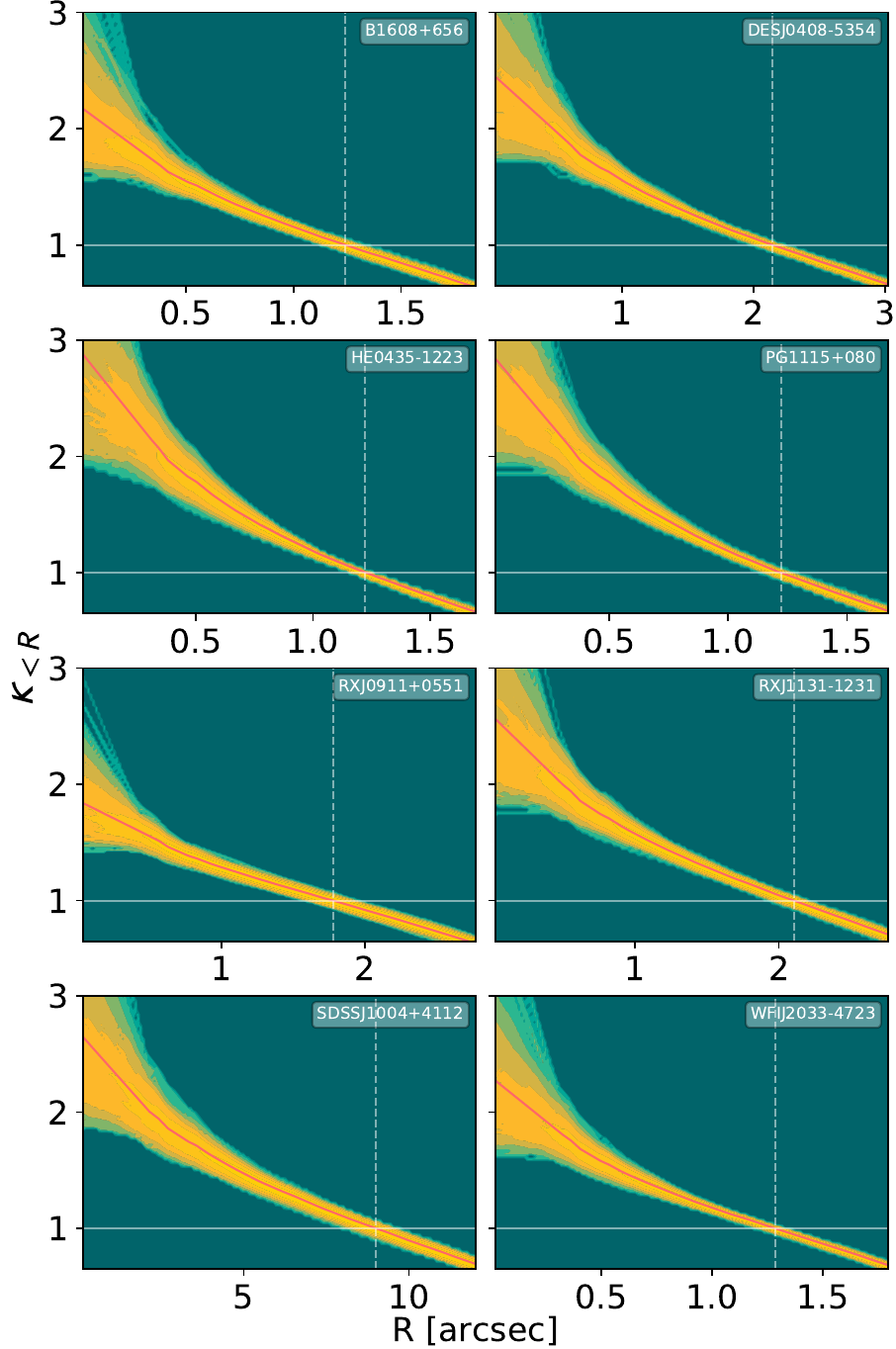}
    \caption{Radial profiles of the mean enclosed $\kappa$ within a given
      projected radius from the centre of the modelled galaxy.  The model
      ensemble is represented with a coloured region with a gradient from green
      to yellow indicating its number density of models. The red line describes
      the ensemble average.  The vertical dashed line shows the notional
      Einstein radius (that is, the radius with mean convergence of unity) for
      the ensemble average.}\figlbl{td:profiles}
\end{figure*}

  \begin{figure*}
    \pcamosaic{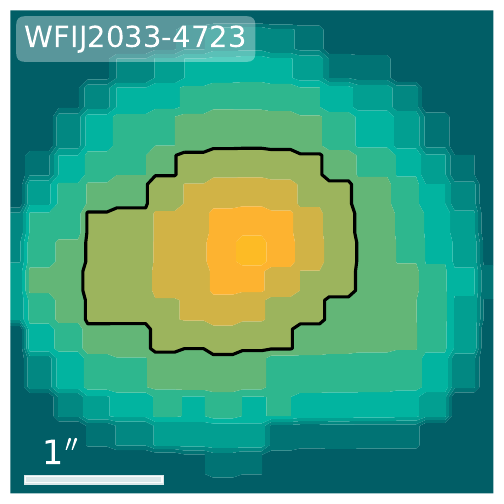}{height=\qheight}
    \caption{Variation of the convergence maps for WFIJ2033-4723 across the
      ensemble.  (The ensemble average appears in the bottom-right panel of
      \protect\figref{td:kappa}.)  The top row shows the 16th and 84th
      percentile projections along the first principal component.  The other rows
      refer to the 2nd, 5th and 100th principal components.  See text near
      \eqref{td:pcaproj} for details.}\figlbl{td:pca}
\end{figure*}

  \begin{figure*}
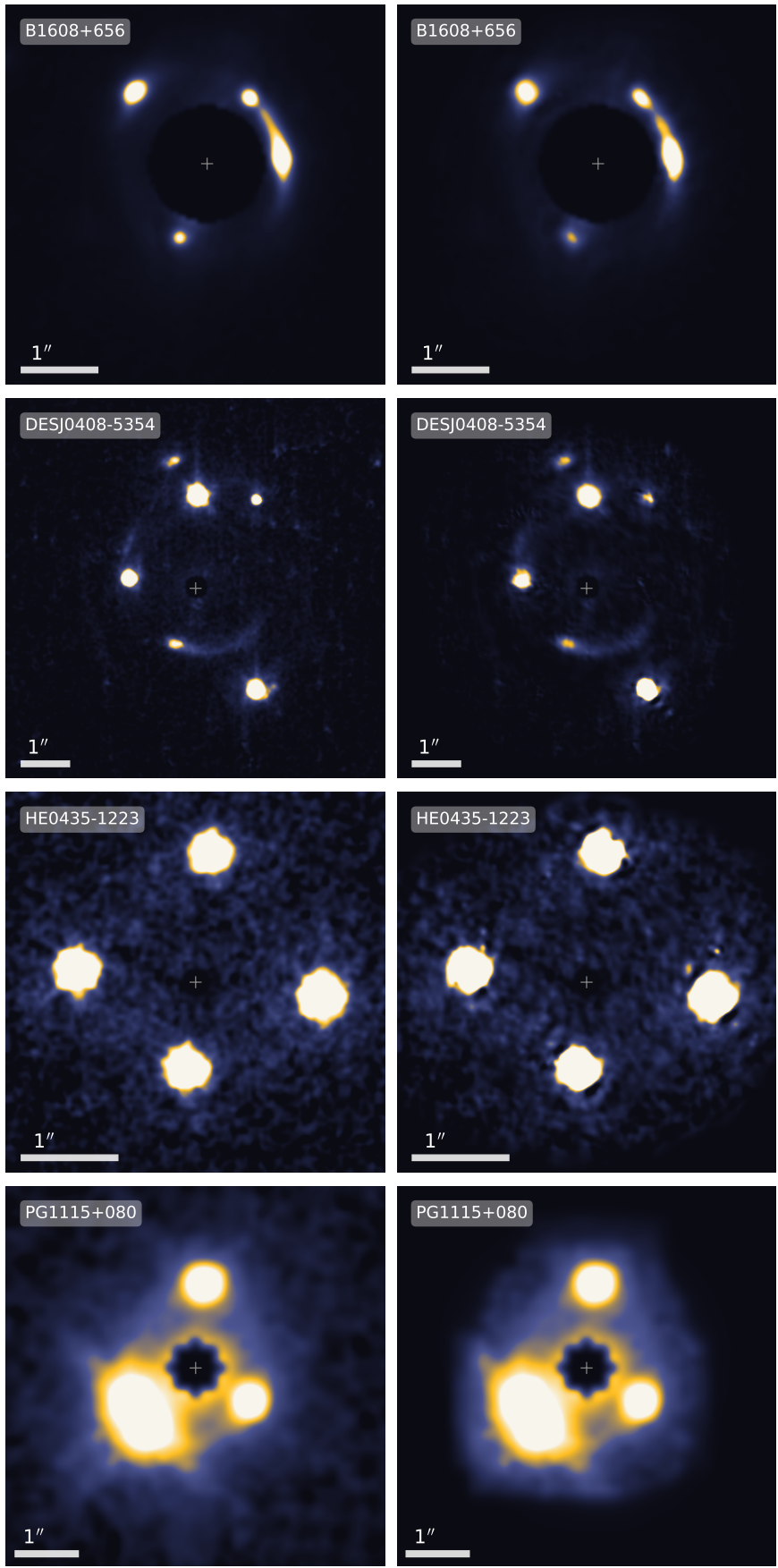

    \centering
    \synthmosaicA{height=\qheight} \caption{Photometric lens data and synthetic
    images produced using the ensemble-average model of each lens.  The maps are
    in arbitrary units of brightness, but for each row adjusted to the same
    brightness levels.  To avoid contamination of the lensed images by the lens
    light, a circular region around the lens has been masked with its center
    indicated by a cross and roughly corresponding to the modelled lens
    position.  The scale bar on the lower left in each panel shows the scale in
    arcseconds.  }\figlbl{td:synthA}
\end{figure*}

\begin{figure*}\ContinuedFloat
    \centering
    \synthmosaicB{height=\qheight}
    \caption{(Continued).}  \figlbl{td:synthB}
\end{figure*}

  \begin{figure}
    \includegraphics[width=\hsize]{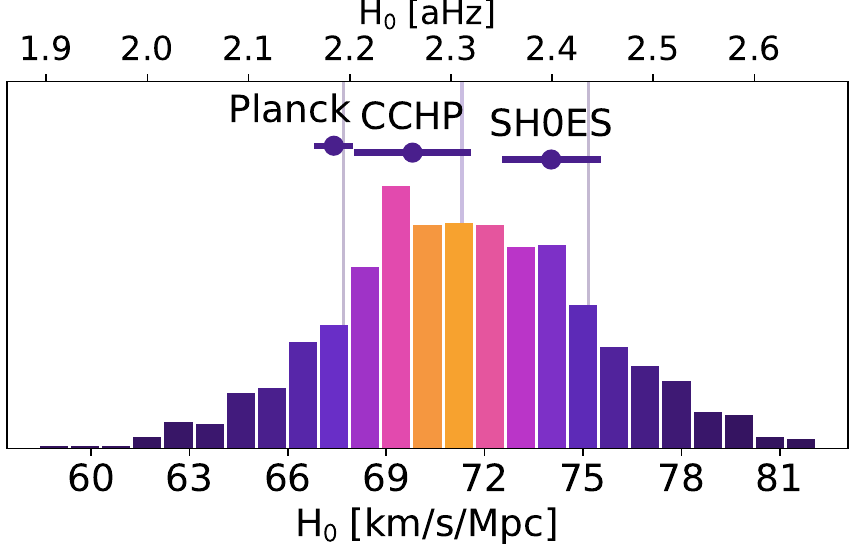}
    \caption{Histogram of the entire ensemble's \Ho{} values.  The ensemble
    consists of 8 simultaneously modelled quad systems.  The vertical lines
    indicate 16th, 50th, and 84th percentiles of the distribution.  To make the
    median furthermore easily discernible, the colouring of the histogram's bars
    corresponds to the cumulative probability centred around the median
    (yellow-magenta-blue-black goes from 1 to 0).  Horizontal error bars
    indicate recent measurements from other methods for comparison: Planck
    \protect\citep{Planck18b}, CCHP\protect\citep[the Carnegie-Chicago Hubble
    Program;][]{Freedman19}, and SH0ES \protect\citep[the Supernovae \Ho{} for
    the Equation of State;][]{Riess19}  }\figlbl{td:H0}
    \includegraphics[width=\hsize]{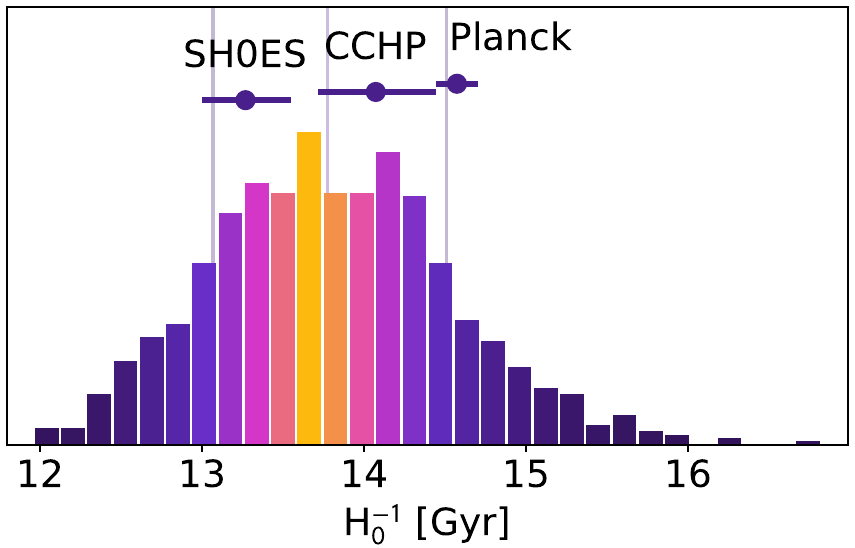}
    \caption{Histogram of the entire ensemble's $\Ho^{-1}$ values, corresponding
    to the distribution in \figref{td:H0}.  }\figlbl{td:invH0}
    \includegraphics[width=\hsize]{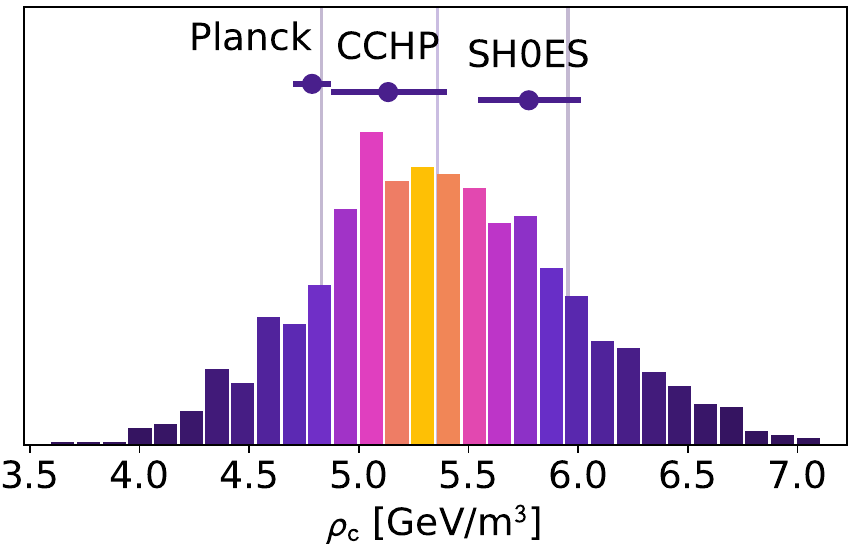}
    \caption{Histogram of the entire ensemble's $\Ho^2$ values in form
      of the cosmological critical density $\rho_{c} = 3/(8\pi
      G)H_{0}^{2}c^2/e$ in $\mathrm{GeV/m^3}$, following
      \protect\figref{td:H0} and \protect\figref*{td:invH0}.  It
      corresponds to an energy density of roughly 1 or 2 alpha
      particles per cubic metre.  }\figlbl{td:rhoH0}
\end{figure}

  \begin{figure}
    \includegraphics[width=0.49\textwidth]{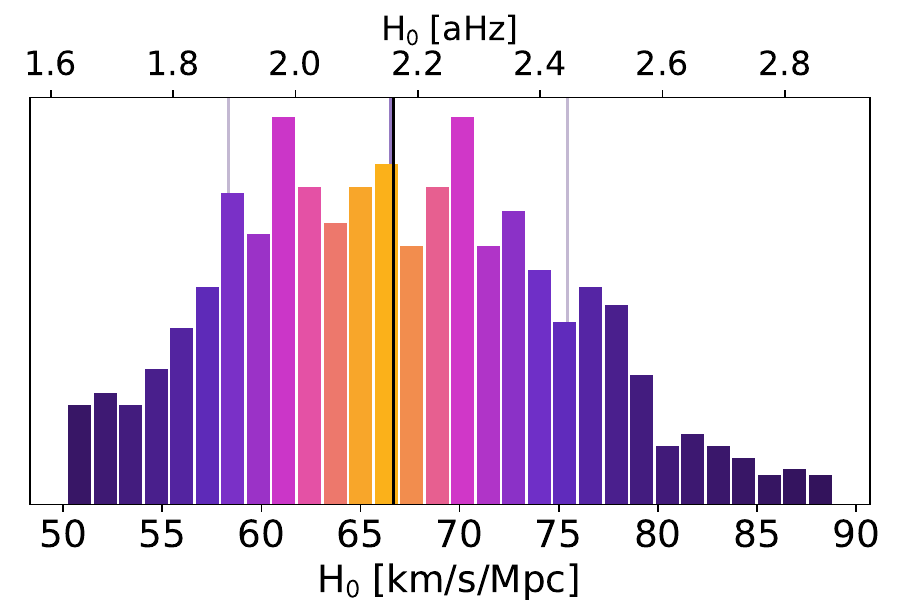}
    \includegraphics[width=0.49\textwidth]{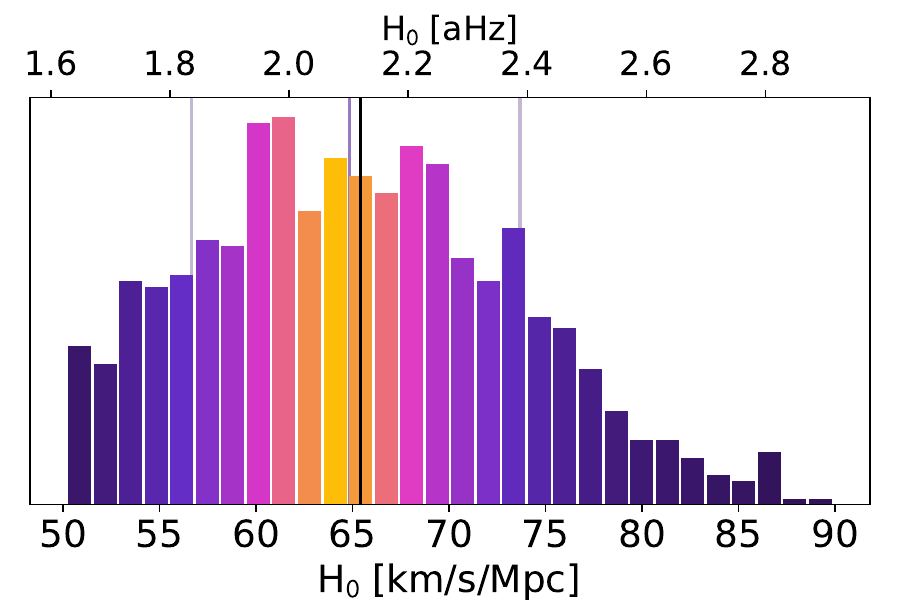}
    \caption{Best results for rung2 (top) and rung3 (bottom) of the TDLMC.
    Generally, we obtained best results from ensembles containing only quads
    which were simultaneously modelled.  The ensembles each consist of 4 quad
    systems from the corresponding rung of the challenge. For these models an
    additional prior was used which required \Ho{} to be higher than 50 and
    lower than 90~\Hunits{}. Red vertical lines indicate the median 68 per-cent
    confidence range of $\Ho{} = 66.5^{+8.9}_{-8.2}\,\Hunits{}$ for rung2 and
    $\Ho{} = 64.9^{+8.8}_{-8.2}\,\Hunits{}$ for rung3.  To make the median
    furthermore easily discernible, the colouring of the histogram bars
    corresponds to the cumulative probability centred around the median
    (yellow-magenta-blue-black goes from 1 to 0).  The black vertical lines
    indicate the truth value for \Ho{} of the corresponding TDLMC rung.  Note
    that these were the \textit{best} results of each rung.  The final
    submission also included models further from the truth, especially when they
    included doubles.}\figlbl{td:tdlmc}
\end{figure}

\label{lastpage}
\clearpage

\end{document}